\documentclass[prb, twocolumn, floatfix, superscriptaddress, longbibliography, reprint]{revtex4-2}
\usepackage[pdftex,plainpages=false,colorlinks=true,linkcolor=blue, citecolor=blue, urlcolor=blue]{hyperref}
\usepackage{amsfonts}
\usepackage{amsmath}
\usepackage{amssymb}
\usepackage{graphicx}
\usepackage{natbib}
\usepackage{color}
\usepackage{placeins}
\usepackage{physics}

\begin{document}
 
\title{Ambipolar doping of a charge-transfer insulator in the Emery model}
\author{G. Sordi}
\email[corresponding author: ]{giovanni.sordi@rhul.ac.uk}
\affiliation{Department of Physics, Royal Holloway, University of London, Egham, Surrey, UK, TW20 0EX}
\author{G. L. Reaney}
\affiliation{Department of Physics, Royal Holloway, University of London, Egham, Surrey, UK, TW20 0EX}
\author{N. Kowalski}
\affiliation{D\'epartement de physique, Institut quantique \& RQMP, Universit\'e de Sherbrooke, Sherbrooke, Qu\'ebec, Canada J1K 2R1}
\author{P. S\'emon}
\affiliation{D\'epartement de physique, Institut quantique \& RQMP, Universit\'e de Sherbrooke, Sherbrooke, Qu\'ebec, Canada J1K 2R1}
\author{A.-M. S. Tremblay}
\affiliation{D\'epartement de physique, Institut quantique \& RQMP, Universit\'e de Sherbrooke, Sherbrooke, Qu\'ebec, Canada J1K 2R1}
\date{\today}

\begin{abstract}
Understanding the similarities and differences between adding or removing electrons from a charge-transfer insulator may provide insights about the origin of the electron-hole asymmetry found in cuprates. 
Here we study with cellular dynamical mean-field theory the Emery model set in the charge-transfer insulator regime, and dope it with either electrons or holes. We consider the normal state only and focus on the doping evolution of the orbital character of the dopants and on the nature of the doping-driven transition. 
Regarding the orbital character of the dopants, we found an electron-hole asymmetry: doped electrons mostly enter the copper orbitals, whereas doped holes mostly enter the oxygen orbitals. 
Regarding the nature of the doping-driven transition, we found no qualitative electron-hole asymmetry: upon either electron or hole doping, there is a two-stage transition from a charge-transfer insulator to a strongly correlated pseudogap and then to a metal. This shows that a strongly correlated pseudogap is an emergent feature of doped correlated insulators in two dimensions, in qualitative agreement with recent experiments on ambipolar Sr$_{1-x}$La$_{x}$CuO$_{2+y}$ cuprate films. 
Our results indicate that merely doping with holes or electrons a charge-transfer insulator is not sufficient for explaining the electron-hole asymmetry observed in the normal state phase diagram of cuprates. 
Our work reinforces the view that actual hole-doped cuprates are more correlated than their electron-doped counterparts. 
\end{abstract}
 
\maketitle

\section{Introduction}
\label{sec:introduction}

The basic building blocks of high temperature superconducting cuprates are the copper-oxygen planes and the layers of ions between them~\cite{keimerRev, Dagotto:RMP1994, ift}. In the copper-oxygen plane, the copper $3d_{x^2-y^2}$ orbital hybridises with the surrounding oxygen $2p_x, 2p_y$ orbitals. The layers give charge neutrality to the system. Hence, by doping these layers, we can add electron to, or remove electrons from (equivalently, add hole to) the copper-oxygen planes. This doping process produces a striking electron-hole asymmetric temperature-doping phase diagram with respect to the zero-doping (“parent”) state~\cite{keimerRev, AMJulich}. However, the origin of this electron-hole asymmetry is still debated~\cite{ift, LeeRMP:2006, Armitage:RMP2010, segawa:NatPhys2010, Cedric:NatPhys2010, Zhong:PRL2020}. 

On the experimental side, one barrier hindering a global view on this electron-hole asymmetry is that this asymmetry is only derived from the comparison of different cuprate materials. In contrast to conventional semiconductors, it has proven to be experimentally difficult to achieve ambipolar doping, i.e. to dope a given “parent” cuprate material with either holes or electrons without altering its crystal structure~\cite{ift}. As a result, one can only indirectly compare hole-doped cuprates, such as La$_{2-x}$Sr$_x$CuO$_4$ (LSCO) and YBa$_2$Cu$_3$O$_y$ (YBCO), to electron-doped cuprates such as Nd$_{2-x}$Ce$_x$CuO$_4$ (NCCO) and Pr$_{2-x}$Ce$_x$CuO$_4$ (PCCO). However, these electron and hole-doped cuprates have a key structural difference, namely the lack of apical oxygens in the electron-doped systems. 
Experimental advances have enabled ambipolar doping only in a few cuprates, including the crystals of  Y$_{1-z}$La$_z$(Ba$_{1-x}$La$_x$)$_2$Cu$_3$O$_y$ (YLBLCO), although for few percent of doping only~\cite{Segawa:PRB2006, segawa:NatPhys2010}, and the cuprate films of Sr$_{1-x}$La$_{x}$CuO$_{2+y}$ (SLCO)~\cite{Zhong:PRL2020}. Remarkably, the high-energy features of the bandstructure of the CuO$_2$ planes of SLCO remain essentially symmetric upon either electron or hole doping~\cite{Zhong:PRL2020}. 

On the theory side, the electron-hole doping asymmetry in cuprates can be rationalised within the Emery model~\cite{Emery_1987}, which in its simplest form includes two oxygen orbitals hybridized with one copper orbital which has onsite electron-electron repulsion. 
This model can be set in different regimes, as classified by the Zaanen-Sawatzky-Allen framework~\cite{zsa}. Within this framework, the relative size of the onsite repulsion $U_d$ and the charge-transfer energy $\Delta$ (i.e., the difference between the energy of the oxygen orbital and the relevant copper orbital) can place the Emery model in a Mott-Hubbard insulating regime (for $U_d<\Delta$), in a charge-transfer insulating regime (for $U_d>\Delta$), in an intermediate regime between the two (for $U_d \approx \Delta$), or in a metallic regime (if the kinetic energy gain overcomes the energy cost to excite an electron across the insulating gap of charge-transfer or Mott-Hubbard type). 

Early calculations~\cite{Kotliar:IJMPB1991, Baumgartel:PRB1993, Scalapino:1994} indicate that real cuprate materials lie close to the boundary between the charge-transfer insulating regime and the metallic regime of the Zaanen-Sawatzky-Allen framework. This suggests two possible mechanisms for the origin of the electron-hole asymmetry observed in cuprates. The first is the location of the dopant carriers, since in a charge-transfer insulator doped holes are expected to occupy the oxygen orbitals and doped electrons to occupy the copper orbital. The second is the relative position of the parent materials relative to the charge-transfer insulator to metal boundary in the Zaanen-Sawatzky-Allen scheme. 
Mean-field calculations in Ref.~\cite{Kotiar:PhysC1988} find an intriguing electron-hole symmetry in the bandstructure of a doped charge-transfer insulator, irrespective of the location of the doped carriers. A combination of density functional theory and dynamical mean-field theory of Refs.~\cite{Cedric:NatPhys2010, cedricApical} suggests that the parent compounds of LSCO and NCCO lie on different sides of this charge-transfer insulator to metal boundary, with LSCO lying on the charge-transfer insulator side and NCCO lying on the correlated metallic side (at low temperature, this metal then becomes insulating due to Slater, and not Mott, mechanism~\cite{Cedric:NatPhys2010, AMJulich, LorenzoAF}). 
\footnote{In the context of the single-band 2D Hubbard model, this translates into modeling hole (electron) doped cuprates by using a value of the interaction strength larger (smaller) than the metal to Mott insulator boundary~\cite{st}. }

\begin{figure}[t!]
\centering{
\includegraphics[width=0.935\linewidth]{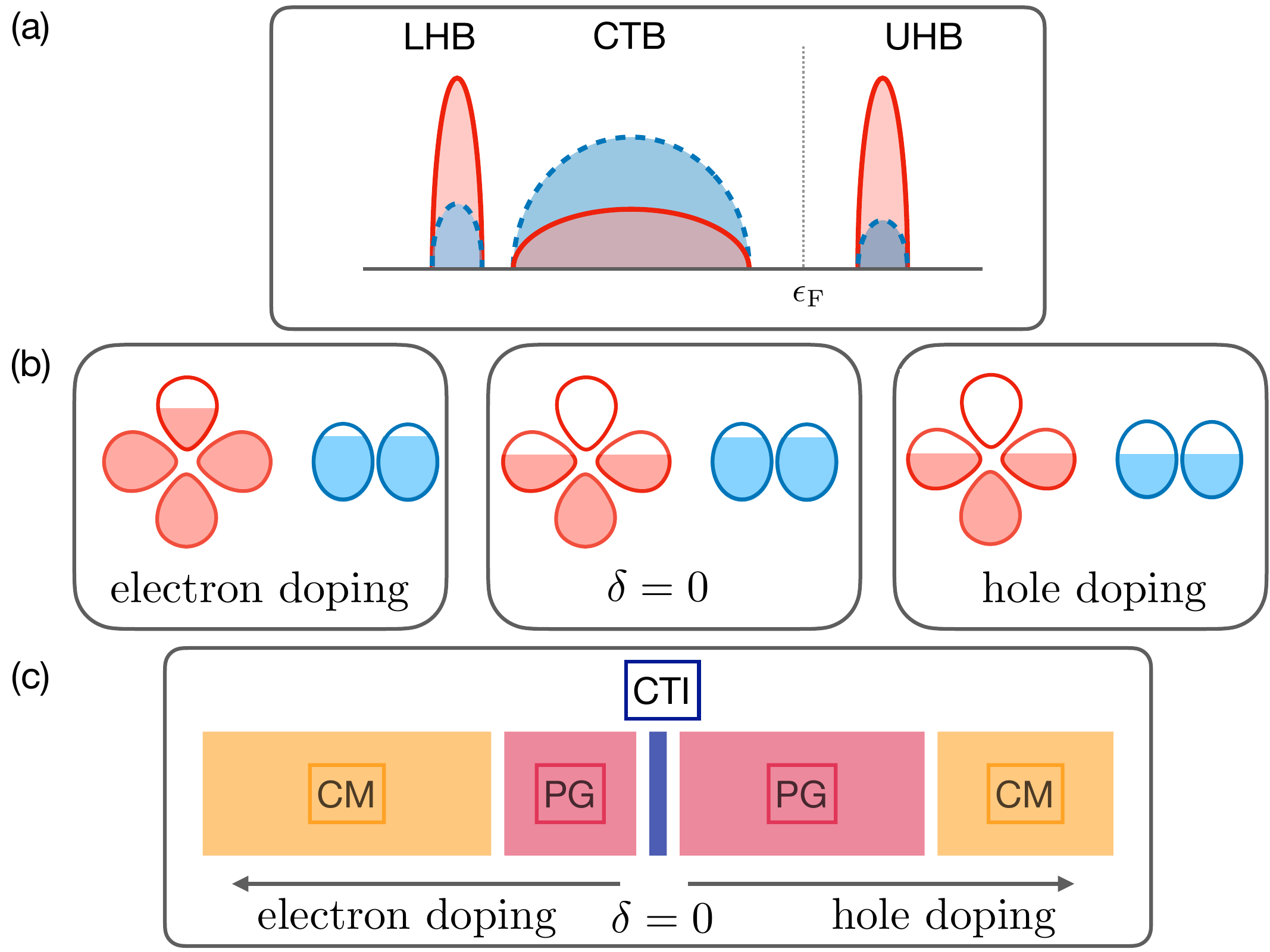}}
\caption{(a) Sketch of the copper $3d$ and oxygen $2p$ partial density of states (red and blue respectively) of an undoped charge-transfer insulator. The charge-transfer band (CTB) occurs between the lower Hubbard band (LHB) and the upper Hubbard band (UHB). The Fermi energy is located between the CTB and the UHB so that the gap is of charge-transfer type. 
(b) Sketch of the $3d$ and $2p$ orbital occupancy at zero doping (middle panel), electron doping (left panel) and hole doping (right panel). The color filling of each orbital indicates the orbital occupancy, whilst the empty space indicates the hole content.
(c) Sketch of the two-stage metal to charge-transfer insulator transition driven by doping. Upon either electron or hole doping, the sequence of the phases is charge-transfer insulator (CTI), pseudogap (PG), and correlated metal (CM). The doping extent of the PG phase is not necessarily symmetric for electron and hole doping.
}
\label{fig:sketch}
\end{figure}

Here, the problem we want to consider is what type of asymmetry results {\it solely} from doping a charge-transfer insulator in the {\it normal state} with either electrons or holes. First, is there any asymmetry in the doping evolution of the orbital character of the dopant carriers upon electron or hole doping? Second, is there any asymmetry in the qualitative features of the normal state doping-driven transition upon electron or hole doping? To address these questions, we consider the finite-temperature normal state of the Emery model set in the charge-transfer insulating regime and dope the resulting charge-transfer insulator with either electron or holes (see sketch of the density of states of the undoped model in Fig.~\ref{fig:sketch}(a)). 

The strong correlations giving rise to the charge-transfer insulator present an outstanding theoretical challenge, necessitating the use of nonperturbative methods. Over the years, significant progress on the physics of the Emery model has been enabled by advanced computational techniques, including quantum Monte Carlo methods~\cite{Dopf:PRB1990, Scalettar:PRB1991, Aoki:PRL1996, Guerrero:PRB1998, Cedric2014, Kung:PRB2016, Vitali:PRB2019, Chiciak:PRB2020}, density matrix renormalization group~\cite{Nishimoto:PRB2002, White:PRB2015, Huang:Science2017}, two-particle self-consistent approach~\cite{Ogura:PRB2015, Chloe:PRB2024}, single-site dynamical mean-field theory~\cite{AntoineCuO2, Lombardo1996, Zolf2000, Cedric2008, DeMedici2009, Cedric:NatPhys2010, cedricApical, Wang:2011, Wang:2011b} and its cluster extensions~\cite{Macridin2005, Kent:PRB2008, Cedric:EPL2012, go, Lorenzo3band, Dash:PRB2019, Mai:PRB2021, Mai:npj2021, Nicolas:PNAS2021}, variational cluster approximation~\cite{ArrigoniCuO2, Hanke:2010}, density matrix embedding theory~\cite{Cui:PRR2020}, and 2D tensor network methods~\cite{Ponsioen:PRB2023}. 
However, due to the complexity of the Emery model, our understanding of the finite-temperature aspects of the normal state doping-driven metal to charge-transfer insulator transition remain incomplete. Our work employs finite-temperature cellular extension~\cite{maier, kotliarRMP, tremblayR} of dynamical mean-field theory~\cite{rmp} (CDMFT) and builds upon Ref.~\cite{Lorenzo3band}, which focused on hole doping only.

The rest of our work is organised as follows. We begin in Section~\ref{sec:findings} by summarising the main findings of our work. Section~\ref{sec:method} briefly presents the Emery model, the cellular dynamical mean-field method used to solve it, and the model parameters of our study. In Section~\ref{sec:occupation} we discuss the metal to charge-transfer insulator transition driven by either electron or hole doping. This allows us to construct and discuss the normal state temperature-doping phase diagram in Section~\ref{sec:phasediagram}. In Section~\ref{sec:dos} we use the density of states to characterise the different phases of the system: a charge-transfer insulating phase at zero doping, a strongly correlated pseudogap phase at small doping, and a more conventional correlated metallic phase at large doping. In Section~\ref{sec:holecontent} we turn to the doping evolution of the orbital character of the dopant carriers. In Section~\ref{sec:Ud} we analyse the effect of the interaction on the copper orbital on the phase diagram and its phases, as well as on the redistribution of the orbital character of the dopant carriers. Section~\ref{sec:discussion} discusses the implications of our findings for real cuprate materials. Finally, in Section~\ref{sec:conclusions} we present our main conclusions.

\section{Summary of key findings}
\label{sec:findings}

In this section we describe the main findings and implications of our work. 
First, we find an asymmetry in the orbital character of the dopant carriers: doped electrons mostly enter the copper orbitals, whereas doped holes mostly enter the oxygen orbitals (see sketch in Fig.~\ref{fig:sketch}(b)). This finding quantifies the expectations for doped charge-transfer insulators~\cite{zsa, Emery:PRB1988, ZhangRice1988} and is compatible with experiments on cuprates~\cite{Tranquada:PRB1987, Bianconi:SSC1987, Nucker:PRB1989, Chen:PRL1991, Chen:PRL1992, Sakurai:Science2011, Jurkutat:PRB2014, Gauquelin2014}. We also find that upon increasing the electron-electron interaction on the copper orbital, the copper hole content increases and the oxygen hole content decreases. This allows us to identify the interaction on the copper orbital as one possible microscopic mechanism controlling the redistribution of electrons and holes in the copper-oxygen plane.
 
Second, we show that the low-temperature normal state doping-driven metal to charge-transfer insulator transition has the {\it same qualitative features} upon either electron or hole doping: at zero doping, there is a continuous charge-transfer insulator to strongly correlated pseudogap transition, and at finite doping there is a discontinuous transition between the strongly correlated pseudogap and a more conventional correlated metal (see sequence of phases in Fig.~\ref{fig:sketch}(c)). The precise value of doping of the pseudogap to metal transition depends on the dopant carriers, but the qualitative features of the normal state doping-driven transition of a charge-transfer insulator is thus {\it symmetric} upon electron or hole doping. This transition is analogous to that found in the doped Mott-Hubbard insulator realised with the single-band 2D Hubbard model~\cite{sht, sht2}, as pointed out in Ref.~\cite{Lorenzo3band} for hole doping.

Our findings that the ambipolar doping of a charge-transfer insulator (i) does produce an asymmetry in the orbital character of the dopants, (ii) but is not sufficient to produce an asymmetry on the qualitative features of the normal state doping-driven transition, enable us to derive two main insights on the microscopic mechanism controlling the doping of a charge-transfer insulator, as we shall discuss in more detail in Section~\ref{sec:discussion}. 

First, when ambipolar doping of a given charge-transfer insulating system is experimentally attainable, a qualitatively electron-hole symmetry in the normal state doping-driven transition should be expected. Indeed, upon ambipolar doping of the cuprates films of SLCO~\cite{Zhong:PRL2020}, a surprising {\it symmetry} in the high-energy features of the bandstructure of the CuO$_2$ planes is found. 
Most importantly, our results indicate that a strongly correlated pseudogap, namely a pseudogap solely arising from Mott physics plus short-range correlations, can emerge from a charge-transfer insulator upon either electron or hole doping, i.e. irrespective of the orbital character of the dopant carriers. This finding is in qualitative agreement with the work on ambipolar doping of cuprate films of SLCO~\cite{Zhong:PRL2020}. Our finding implies that a strongly correlated pseudogap is an emergent feature of doped correlated insulators in two dimensions.  

Second, when ambipolar doping is not experimentally attainable and an indirect comparison between hole-doped and electron-doped cuprates is performed (say, between LSCO and NCCO, for concreteness), our results rule out that the source of the observed electron-hole asymmetry in the normal state lies solely in doping with holes or electrons the same parent charge-transfer insulator. However, our results cannot directly identify the source of that asymmetry. 
On one hand, we can empirically observe that the basic phenomenology of the normal state of real hole-doped cuprates such as LSCO is well captured by that of a hole-doped charge-transfer insulator and that the basic phenomenology of the normal state of real electron-doped cuprates such as NCCO is not captured by that of an electron-doped charge-transfer insulator. 
On the other hand, we show that the large oxygen hole content found in experiments on electron doped cuprates~\cite{Jurkutat:PRB2014} is inconsistent with the set of parameters used in our work and correlates with a small value of the electron-electron interaction on the copper orbital. 
These findings are compatible with the view that actual hole-doped cuprates are more correlated than their electron-doped counterparts~\cite{st, Cedric:NatPhys2010, cedricApical}.

\section{Model and method}
\label{sec:method}

\subsection{Emery model}

\begin{figure}[ht!]
\centering{
\includegraphics[width=1.\linewidth]{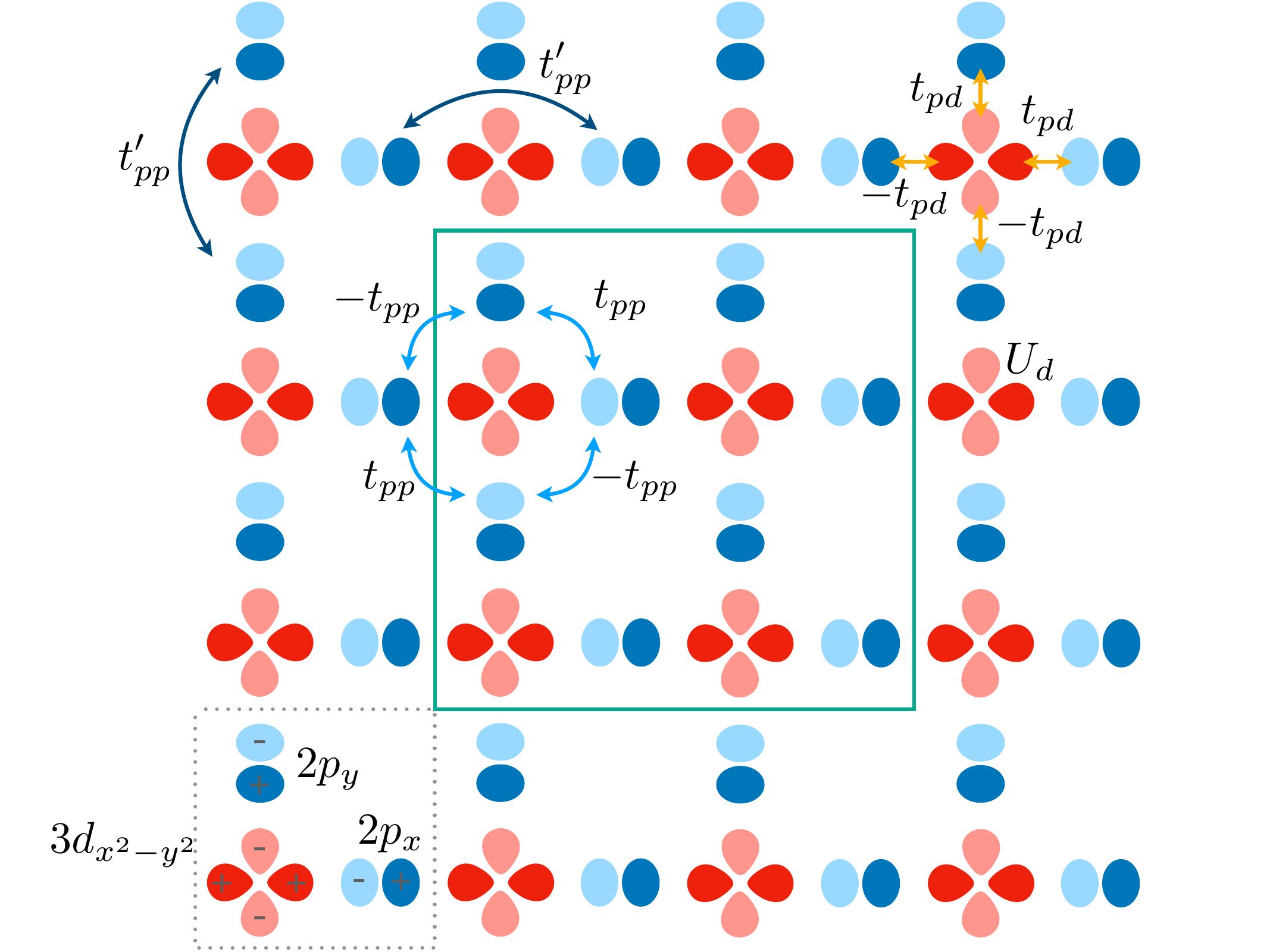}}
\caption{Sketch of the Emery model. The dotted grey square illustrates the unit cell. The copper $3d$  orbitals are shown in red and the oxygen $2p_x$, $2p_y$ orbitals are shown in blue. Bright (pale) colours indicate the positive (negative) sign of the orbital lobes. In our convention, the overlap between orbitals of the opposite (equal) sign gives rise to a positive (negative) hopping term. $t_{pd}$ is the hopping amplitude between the nearest-neighbor Cu-O orbitals (orange arrows). $t_{pp}$ is the hopping amplitude between the nearest-neighbour $2p_x$ and $2p_y$ orbitals (light blue arrows). $t_{pp}^\prime$ is the hopping amplitude between two $2p_x$ or two $2p_y$ orbitals, across the Cu orbital (dark blue arrows). $U_d$ is the onsite electron-electron repulsion on the Cu orbital. The solid green square illustrates the cluster we used in CDMFT. 
}
\label{fig:model}
\end{figure}

The Emery model considers the copper $3d_{x^2-y^2}$ and oxygen $2p_x, 2p_y$ orbitals in the CuO$_2$ plane~\cite{Emery_1987}. The lattice structure is shown in Fig.~\ref{fig:model}. We take the square grey dotted line as our unit cell, which includes one copper $3d_{x^2-y^2}$ orbital (schematically represented by a red boundary surface with four lobes) and the oxygen $2p_x, 2p_y$ orbitals (schematically represented by a blue boundary surface with two lobes). Let $\mathbf{R}_i$ be the position of the Cu $3d_{x^2-y^2}$ orbital and $\mathbf{a}_x, \mathbf{a}_y$ be the Bravais lattice vectors, whose length is equal to the nearest neighbors Cu-Cu distance. Let us assume $\mathbf{a}_x = \mathbf{a}_y$, so that the Cu $3d_{x^2-y^2}$ orbitals form a square lattice. The $2p_x, 2p_y$ orbitals are placed at the middle of the nearest neighbors Cu-Cu distance, so that they form a square lattice rotated by $\pi/4$ relative to the Cu lattice. Let $\mathbf{R}_i +\mathbf{r}_x$ be the position of the oxygen 2$p_x$ orbital inside the unit cell and $\mathbf{R}_i +\mathbf{r}_y$ be the position of the oxygen 2$p_y$ orbital inside the unit cell. 

The Emery model takes into account three hopping processes. An electron can hop from an oxygen orbital (i) to a nearest-neighbor copper orbital via the hopping amplitude $t_{pd}$ (orange arrows), (ii) to a nearest-neighbor (diagonal) oxygen orbital via the hopping amplitude $t_{pp}$ (light blue arrows), and (iii) to an oxygen orbital across the copper orbital  via the hopping amplitude $t_{pp}^\prime$ (dark blue arrows).
The sign of the copper-oxygen and oxygen-oxygen hopping follows the convention indicated in Fig.~\ref{fig:model}. 
Let $\epsilon_d$ be the energy of the copper $3d_{x^2-y^2}$ orbitals  and $\epsilon_p$ be the energy of the oxygen $2p_x, 2p_y$ orbitals. 
Regarding the interaction processes, on a copper orbital two electrons experience a Coulomb repulsion $U_d$. For simplicity, we neglect the Coulomb repulsion $U_p$ between electrons on an oxygen orbital. 

To write the Emery model hamiltonian, let us introduce the creation and destruction operators of electrons at different lattice sites. Let $d_{\mathbf{R}_{i} \sigma}^\dagger$ ($d_{\mathbf{R}_{i} \sigma}$) be the operator that creates (destroys) an electron with spin $\sigma$ at the copper orbital at site $\mathbf{R}_{i}$ and $n_{d \mathbf{R}_{i} \sigma} = d_{\mathbf{R}_{i} \sigma}^\dagger d_{\mathbf{R}_{i} \sigma}$ the associated number operator. 
Similarly, let $p_{x \mathbf{R}_{i} +\mathbf{r}_{x} \sigma}^{\dagger}$ ($p_{x \mathbf{R}_{i} +\mathbf{r}_{x} \sigma}$) be the operator that creates (destroys) an electron with spin $\sigma$ at the $2p_x$ oxygen orbital at site $\mathbf{R}_{i}+\mathbf{r}_{x}$. Likewise, $p_{y \mathbf{R}_{i} +\mathbf{r}_{y} \sigma}^{\dagger}$ ($p_{y \mathbf{R}_{i} +\mathbf{r}_{y} \sigma}$) acts on the $2p_y$ oxygen orbital at site $\mathbf{R}_{i}+\mathbf{r}_{y}$.

Thus, in real space the Emery model can be written as:
\begin{align}
& H  =  
\left( \epsilon_{d} -\mu \right) \sum_{\mathbf{R}_{i} \sigma} d_{\mathbf{R}_{i} \sigma}^{\dagger} d_{\mathbf{R}_{i} \sigma}
\nonumber \\
& + \left( \epsilon_{p} -\mu \right)  \sum_{\mathbf{R}_{i} \sigma} \left( p_{x \mathbf{R}_{i} +\mathbf{r}_{x} \sigma}^{\dagger} p_{ x \mathbf{R}_{i} +\mathbf{r}_{x} \sigma} + p_{y\mathbf{R}_{i} +\mathbf{r}_{y} \sigma}^{\dagger} p_{y \mathbf{R}_{i} +\mathbf{r}_{y} \sigma}  \right)
\nonumber \\
& + U_d \sum_{\mathbf{R}_{i}} n_{d \mathbf{R}_{i} \uparrow} n_{d \mathbf{R}_{i} \downarrow} \nonumber \\
& +t_{pd} \sum_{\mathbf{R}_{i} \sigma} \left[  \left( d_{\mathbf{R}_{i} \sigma}^{\dagger} p_{x \mathbf{R}_{i} +\mathbf{r}_{x} \sigma} 
+ d_{\mathbf{R}_{i} \sigma}^{\dagger} p_{y \mathbf{R}_{i} +\mathbf{r}_{y} \sigma}  \right. \right. \nonumber \\
& \left. \left. \qquad - p_{x \mathbf{R}_{i} +\mathbf{r}_{x} \sigma}^{\dagger}d_{\mathbf{R}_{i} +\mathbf{a}_{x} \sigma} 
- p_{y \mathbf{R}_{i} +\mathbf{r}_{y} \sigma}^{\dagger}d_{\mathbf{R}_{i} +\mathbf{a}_{y} \sigma} \right) +\textrm{h.c.} \right]  \nonumber \\
& +t_{pp}^\prime \sum_{\mathbf{R}_{i} \sigma} \left[ \left( p_{x \mathbf{R}_{i} +\mathbf{r}_{x} \sigma}^{\dagger} p_{x \mathbf{R}_{i} +\mathbf{a}_{x} +\mathbf{r}_{x} \sigma} \right. \right. \nonumber \\
& \left. \left. \qquad + p_{y \mathbf{R}_{i} +\mathbf{r}_{y} \sigma}^{\dagger} p_{y \mathbf{R}_{i} +\mathbf{a}_{y} +\mathbf{r}_{y} \sigma} \right)  +\textrm{h.c.} \right]  \nonumber \\
& +t_{pp} \sum_{\mathbf{R}_{i} \sigma} \left\{ \left[ \left( -p_{x \mathbf{R}_{i} +\mathbf{r}_{x} \sigma}^{\dagger} p_{y \mathbf{R}_{i} +\mathbf{a}_{x} +\mathbf{r}_{y} \sigma} \right. \right. \right. \nonumber \\
& \left. \left. \left. \qquad + p_{x \mathbf{R}_{i} +\mathbf{r}_{x} \sigma}^{\dagger} p_{y \mathbf{R}_{i} +\mathbf{a}_{x} -\mathbf{a}_{y} +\mathbf{r}_{y} \sigma} \right) +\textrm{h.c.} \right] \right. \nonumber \\
&   \qquad \; + \left. \left[ \left( -p_{y \mathbf{R}_{i} +\mathbf{r}_{y} \sigma}^{\dagger} p_{x \mathbf{R}_{i} +\mathbf{a}_{y} +\mathbf{r}_{x} \sigma} \right. \right. \right. \nonumber \\
& \left. \left. \left. \qquad + p_{y \mathbf{R}_{i} +\mathbf{r}_{y} \sigma}^{\dagger} p_{x \mathbf{R}_{i} +\mathbf{r}_{x} \sigma} \right)  +\textrm{h.c.} \right] \right\} .
\label{eq:EmeryModel_realspace}
\end{align}

Here, the chemical potential $\mu$ controls the filling of the system. In writing Eq.~\ref{eq:EmeryModel_realspace}, we consider all hopping terms with the creation operators in the unit cell and the destruction operators with whom they are related, either by hopping within the same unit cell or from the adjacent unit cells, so to pave the lattice without double counting. The hermitian conjugate ($h.c.$) terms account for the reverse hopping directions. 
 
It is convenient to rewrite the noninteracting part of the Emery model Eq.~\ref{eq:EmeryModel_realspace} in reciprocal space. To do that, let us introduce the Fourier transforms
\begin{equation}
\begin{cases}
d_{\mathbf{k} \sigma} = \frac{1}{\sqrt{N}} \sum \limits_{\mathbf{R}_{i}} e^{-i\mathbf{k\cdot R}_{i}} d_{\mathbf{R}_{i} \sigma} , \\ 
d_{\mathbf{R}_{i} \sigma} = \frac{1}{\sqrt{N}} \sum \limits_{\mathbf{k}} e^{i\mathbf{k\cdot R}_{i}} d_{\mathbf{k} \sigma} , \\
\end{cases}
\end{equation}
\begin{equation}
\begin{cases}
  p_{x \mathbf{k} \sigma} = \frac{1}{\sqrt{N}} \sum \limits_{\mathbf{R}_{i}} e^{-i\mathbf{k\cdot R}_{i}} p_{x \mathbf{R}_{i} +\mathbf{r}_{x} \sigma}, \\
p_{x \mathbf{R}_{i} +\mathbf{r}_{x} \sigma}  = \frac{1}{\sqrt{N}} \sum \limits_{\mathbf{k}} e^{i\mathbf{k\cdot R}_{i}} p_{x \mathbf{k} \sigma} ,  \\
\end{cases}
\end{equation}
\begin{equation}
\begin{cases}
 p_{y \mathbf{k} \sigma} = \frac{1}{\sqrt{N}} \sum \limits_{\mathbf{R}_{i}} e^{-i\mathbf{k\cdot R}_{i}} p_{y \mathbf{R}_{i} +\mathbf{r}_{y} \sigma} , \\
 p_{y \mathbf{R}_{i} +\mathbf{r}_{y} \sigma} = \frac{1}{\sqrt{N}} \sum \limits_{\mathbf{k}} e^{i\mathbf{k\cdot R}_{i}} p_{y \mathbf{k} \sigma}  .
\end{cases}
\end{equation}
Note that we use the same phases for all atoms in a given unit cell. 
By ordering the operators in the vectors $C_{\mathbf{k} \sigma} = \left( d_{\mathbf{k} \sigma},  p_{x \mathbf{k} \sigma}, p_{y \mathbf{k} \sigma} \right)^{T}$ and $C_{\mathbf{k} \sigma}^{\dagger} = \left( d_{\mathbf{k} \sigma}^{\dagger} ,  p_{x \mathbf{k}  \sigma}^{\dagger} , p_{y \mathbf{k} \sigma}^{\dagger}  \right)$, the Emery model hamiltonian can be written as
\begin{align}
H & = \sum_{\mathbf{k} \sigma} C_{\mathbf{k} \sigma}^{\dagger}  \left( \mathbf{h}_{0}(\mathbf{k}) -\mu \mathbf{I} \right)  C_{\mathbf{k} \sigma} 
+ U_d \sum_{\mathbf{R}_{i}} n_{d \mathbf{R}_{i} \uparrow} n_{d \mathbf{R}_{i} \downarrow} ,  \nonumber \\
\end{align}
with
\begin{widetext}
\begin{align}
\mathbf{h}_0 (\mathbf{k}) & =
\left(
\begin{array}
[c]{ccc}
\epsilon_{d} & t_{pd}\left(  1-e^{-ik_{x}}\right)   & t_{pd}\left(1-e^{-ik_{y}}\right)  \\
t_{pd}\left(  1-e^{ik_{x}}\right)   & \epsilon_{p} -2t_{pp} +2t_{pp}^\prime \cos k_{x} & t_{pp}\left(  1-e^{ik_{x}}\right)  \left(  1-e^{-ik_{y}}\right)  \\
t_{pd}\left(  1-e^{ik_{y}}\right)   & t_{pp}\left(  1-e^{-ik_{x}}\right) \left(  1-e^{ik_{y}}\right)   & \epsilon_{p}-2t_{pp}+2t_{pp}^\prime \cos k_{y}
\end{array} \right)  , 
\label{eq:h0}
\end{align}
\end{widetext}
where we have set $\mathbf{a}_x = \mathbf{a}_y= 1$. Following the convention of Ref.~\cite{AndersenLDA}, in Eq.~\ref{eq:h0} we have also renormalized the onsite oxygen energy by replacing $\epsilon_p$ with $\epsilon_p -2t_{pp}$.

\subsection{Cellular dynamical mean-field theory}

To solve the Emery model Eq.~\ref{eq:EmeryModel_realspace} at finite temperature, we use the cellular extension~\cite{maier, kotliarRMP, tremblayR} of dynamical mean-field theory~\cite{rmp} (CDMFT) and follow the procedure described in Ref.~\cite{Lorenzo3band}. Here we consider the CDMFT equations in the normal state only. The basic idea of CDMFT is to partition the lattice into a superlattice of clusters, to single out one cluster from the lattice and embed it in a self-consistent reservoir of noninteracting electrons. In this work we consider a cluster of $12$ sites with $N_d=4$ copper sites and  $N_p=8$ oxygen sites, as indicated by the green square in Fig.~\ref{fig:model}. 

The electron-electron interaction $U_d$ is on the copper orbitals only. Hence it is useful to integrate out the oxygen orbitals before applying the CDMFT procedure. By integrating out the oxygen orbitals, one obtains an effective single-band model. The noninteracting Green's function of this effective single-band model is 
\begin{align}
G_{0 \, \rm{eff}} (i\omega_n, \mathbf{k})  & = \left[ (i\omega_n +\mu) \mathbf{I} - \mathbf{h}_{0} (\mathbf{k}) \right]_{[dd]}^{-1} , 
\label{eq:G0eff}
\end{align}
where $i\omega_n$ are Matsubara frequencies. 

One can then apply the CDMFT procedure for this effective single-band model. The action of the cluster quantum impurity (cluster of the effective model of $N_d=4$ copper orbitals plus bath) is 
\begin{align}
S = & - \int_{0}^{\beta} d\tau \int_{0}^{\beta} d\tau^\prime \, \boldsymbol{\hat{\psi}}^\dagger(\tau) \boldsymbol{\mathcal{G}_{0}}^{-1} (\tau, \tau^\prime) \boldsymbol{\hat{\psi}}(\tau^\prime) \nonumber \\ 
& + U_d \int_{0}^{\beta} d\tau \, \mathbf{\hat{n}}_{\mathbf{R_i} \uparrow}(\tau) \mathbf{\hat{n}}_{\mathbf{R_i} \downarrow}(\tau)  .
\label{eq:Simp}
\end{align}
Here $\boldsymbol{\hat{\psi}}^\dagger= (\hat{d}_{1 \uparrow}^\dagger \, \cdots \, \hat{d}_{N_d \uparrow}^\dagger \, \hat{d}_{1 \downarrow}^\dagger \, \cdots \, \hat{d}_{N_d \downarrow}^\dagger)$ and $\boldsymbol{\hat{\psi}} = (\hat{d}_{1 \uparrow} \, \cdots \, \hat{d}_{N_d \uparrow} \, \hat{d}_{1 \downarrow} \, \cdots \, \hat{d}_{N_d \downarrow})^T$ are vectors of the Grassmann variables $\hat{d}_{\mathbf{R}_{i} \sigma}^\dagger, \hat{d}_{\mathbf{R}_{i} \sigma}$ corresponding to the operators acting on the copper sites of the cluster and $\boldsymbol{\cal G}_0$ is the Green's function of the noninteracting impurity defined as
\begin{align}
\boldsymbol{\cal G}_0^{-1} (i\omega_n) & = (i\omega_n +\mu) \mathbf{I} - \mathbf{t}_{\rm cl} - \boldsymbol{\Delta}(i\omega_n) .
\label{eq:WeissField}
\end{align}
Here the hybridization matrix function $\boldsymbol{\Delta}(i\omega_n)$ encodes the amplitude processes of the electrons hopping from the cluster to the bath and then back to the cluster. 
The cluster hopping matrix $\mathbf{t_{\rm cl}}$ is defined as ${\bf t_{\rm cl}} = \int d \mathbf{\tilde{k}} \, {\mathbf t}(\mathbf{\tilde{k}})$,  where ${\mathbf t}(\mathbf{\tilde{k}})$ is the lattice hopping matrix in the supercell notation and $\mathbf{\tilde{k}}$ runs over the reduced Brillouin zone of the superlattice. Since here there is no direct hopping between electrons on the copper sites, $\mathbf{t}_{\rm cl}$ is diagonal. 

For a given $\boldsymbol{\cal G}_0$ (or, equivalently, $\boldsymbol{\Delta}$), the solution of the quantum impurity model Eq.~\ref{eq:Simp} gives the cluster Green's function
\begin{align}
\mathbf{G}_{\rm cl} (\tau-\tau^\prime) & = - \langle T_\tau \, \boldsymbol{\hat{\psi}}(\tau) \boldsymbol{\hat{\psi}}^\dagger (\tau') \rangle_{S} . 
\label{eq:G_cl}
\end{align}
From the Dyson equation, the cluster self-energy is
\begin{align}
\boldsymbol{\Sigma}_{\rm cl} (i\omega_n) & = \boldsymbol{\mathcal{G}}_{0}^{-1} (i\omega_n) - \mathbf{G}_{\rm cl}^{-1} (i\omega_n) . 
\label{eq:Sigma_cl}
\end{align}

To fix $\boldsymbol{\mathcal{G}}_{0}$, a self-consistency condition is needed. The self-consistency condition requires that the cluster Green's function $\mathbf{G}_{\rm cl}$ coincides with the superlattice averaged Green's function $\mathbf{\bar{G}}$, which is defined as
\begin{align}
\mathbf{\bar{G}} (i\omega_n)  & = \frac{N_d}{(2\pi)^2} \int d \mathbf{\tilde{k}} \, \left[ (i\omega_n +\mu){\mathbf I} -{\mathbf t}(\mathbf{\tilde{k}}) -\boldsymbol{\Sigma}_{\rm latt} (i\omega_n, \mathbf{\tilde{k}} ) \right]^{-1} .
\end{align}
To identify ${\mathbf G}_{\rm cl}$ with $\boldsymbol{\bar{G}}$, we approximate $\boldsymbol{\Sigma}_{\rm latt} (i\omega_n, \mathbf{\tilde{k}})$ with $\boldsymbol{\Sigma}_{\rm cl} (i\omega_n)$, i.e.
\begin{align}
\mathbf{\bar{G}} (i\omega_n) & \approx \frac{N_d}{(2\pi)^2} \int d \mathbf{\tilde{k}} \, \left[ (i\omega_n +\mu){\bf I} -{\bf t}(\mathbf{\tilde{k}}) -\boldsymbol{\Sigma}_{\rm cl} (i\omega_n) \right]^{-1} .
\label{eq:Gbar}
\end{align}
Hence, in terms of $\boldsymbol{\mathcal{G}}_{0}$, the self-consistency condition can be written as
\begin{align}
& \left( \boldsymbol{\mathcal{G}}_{0}^{-1}(i\omega_n)  - \boldsymbol{\Sigma}_{\rm cl}(i\omega_n)  \right)^{-1} \nonumber \\
& = \frac{N_d}{(2\pi)^2} \int d\mathbf{\tilde{k}} \,  \left[ \mathbf{G}_{0 \, {\rm eff}}^{-1} (i\omega_n, \mathbf{\tilde{k}}) -\boldsymbol{\Sigma}_{\rm cl} (i\omega_n)  \right]^{-1} , 
\end{align}
where  
\begin{align}
& \left[ \mathbf{G}_{0 \, {\rm eff}}^{-1} (i\omega_n, \mathbf{\tilde{k}}) \right]_{\mathbf{R_{i} R_{j}}} \nonumber \\ 
& = \sum_{\mathbf{K}} e^{i \left( \mathbf{K +\tilde{k}} \right) \cdot \left( \mathbf{R_{i} -R_{j} } \right)}  G_{0 \, {\rm eff}}^{-1}  (i\omega_n, \mathbf{K +\tilde{k} } )
\end{align} 
is the effective noninteracting lattice Green's function $G_{0 \, {\rm eff}}$ defined in Eq.~\ref{eq:G0eff} written in the mixed real/momentum space representation~\cite{maier}. Recall that the positions $\left\{ \mathbf{R_{i}} \right\}$ are on the cluster. 

We solve the cluster quantum impurity model Eq.~\ref{eq:Simp} using the hybridization expansion continuous-time quantum Monte Carlo method (CT-HYB)~\cite{millisRMP, Werner:2006, hauleCTQMC, patrickSkipList}. Since the there is no direct hopping process between electrons on the Cu orbitals, we can use the efficient ``segment'' representation~\cite{millisRMP} of the hybridization expansion, as described in Refs.~\cite{Lorenzo3band, Nicolas:PNAS2021}.

\subsection{Model parameters of this work}

The focus of this work is on the electron and hole doping of a charge-transfer insulator. To set the Emery model in the charge-transfer insulating regime, we consider the following set of parameters: 
\begin{align}
& \epsilon_d =0, \epsilon_p =8, t_{pp}=1, t_{pp}^\prime=1, t_{pd}=1.5 . 
\label{eq:modelparameters}
\end{align}
In our work we set $t_{pp}=1$ as our energy unit. 

The rationale for this choice of parameters is as follows: for $U_d=0$ and $t_{pd}=0$, the choice $\epsilon_d =0, \epsilon_p =8$ places the $3d$ level derived from the Cu $3d$ orbitals just below the oxygen bands formed by the overlap of the O $2p$ orbitals. A value of the interaction strength $U_d$ larger than the bare charge-transfer energy $|\epsilon_p-\epsilon_d|$ splits the Cu $3d$ level into two levels of energy $\epsilon_d$ and $\epsilon_d+U_d$, with the latter lying above the oxygen band, hence creating a charge-transfer gap between the O $2p$ bands and the relevant Cu $3d$ level~\cite{zsa}. The charge-transfer insulator thus occurs for one hole per CuO$_2$ unit cell, or equivalently for a total occupation $n_{\rm tot}=n_d +2n_p=5$. 

A finite $t_{pd}$ then has two main effects: it turns the two $3d$ levels split by $U_d$ into bands, namely the lower and upper Hubbard bands, and mixes the character of the Cu $3d$ and O $2p$ bands. The charge-transfer gap thus occurs between a band with mostly $p$ character --the so called charge-transfer band -- and the upper Hubbard band with mostly $d$ character (see sketch of the density of state in Fig.~\ref{fig:sketch}(a)). 

In the Zaanen-Sawatzky-Allen framework~\cite{zsa}, our choice of parameters Eq.~\ref{eq:modelparameters} sets the Emery model in such a ``charge-transfer insulator'' regime. In Sections~\ref{sec:occupation}-\ref{sec:holecontent} we fix the interaction strength at $U_d=11.1$, a value large enough to open a charge-transfer gap at $n_{\rm tot}=5$. Then in Section~\ref{sec:Ud} we vary $U_d$ in the range $U_d \in  \left[ 10, 15 \right]$, which allows us to tune the model across the metal to charge-transfer transition at $n_{\rm tot}=5$.

Let us also comment our choice of $|\epsilon_p-\epsilon_d|=8$. This choice follows from the goal of our work to investigate the effect of doping a clear charge-transfer insulator. A smaller value of $|\epsilon_p-\epsilon_d|$ would be arguably closer to the values obtained in bandstructure calculations for cuprates~\cite{AndersenLDA, Cedric:NatPhys2010}. However, this would increase the mixed character between the $3d$ and $2p$ orbitals, which, in the Zaanen-Sawatzky-Allen framework~\cite{zsa}, moves the Emery model towards the ``intermediate'' (or covalent or mixed-valence) region. In our work we wish to set the model in a clear charge-transfer insulator regime, that is, away from the intermediate regime. In the language of Ref.~\cite{Nicolas:PNAS2021}, our choice of parameters corresponds to an ``ionic charge-transfer insulator'', and a smaller value of $|\epsilon_p-\epsilon_d|$ corresponds to a ``covalent charge-transfer insulator''. 
Furthermore, small values of $|\epsilon_p-\epsilon_d|$ drastically reduce the Monte Carlo sign, leading to a more severe sign problem.

The set of parameters Eq.~\ref{eq:modelparameters} follows those used in Ref.~\cite{Lorenzo3band}, apart from $\epsilon_p$. In Ref.~\cite{Lorenzo3band}, $\epsilon_p=9$, whereas here we use a slightly smaller value, $\epsilon_p=8$. 
Our work further extends Ref.~\cite{Lorenzo3band} in two main directions. First, it confirms the results of Ref.~\cite{Lorenzo3band}  upon hole doping for a slightly smaller value of $\epsilon_p$, which increases the mixed character between the $3d$ and $2p$ orbitals, but still keeps the Emery model in the `ionic' charge-transfer insulator regime. Second, and most importantly, our work studies the effect of both electron and hole doping: this allows us to investigate the electron-hole asymmetry due to doping a band with mostly $p$ character (the charge-transfer band, upon hole doping) and doping a band with mostly $d$ character (the upper Hubbard band, upon electron doping).

\section{Metal to charge-transfer insulator transition driven by doping}
\label{sec:occupation}

This section focuses on the metal to charge-transfer insulator transition induced by either hole or electron doping. To do that, we study the behavior of the doping $\delta$ as a function of the chemical potential $\mu$.  This allows us to reveal the two-stage character of the metal to charge-transfer insulator transition, as well as a peak in the thermodynamic entropy and in the charge compressibility at finite doping levels. 

\subsection{Two-stage metal to charge-transfer insulator transition}
\begin{figure}[ht!]
\centering{
\includegraphics[width=1.\linewidth]{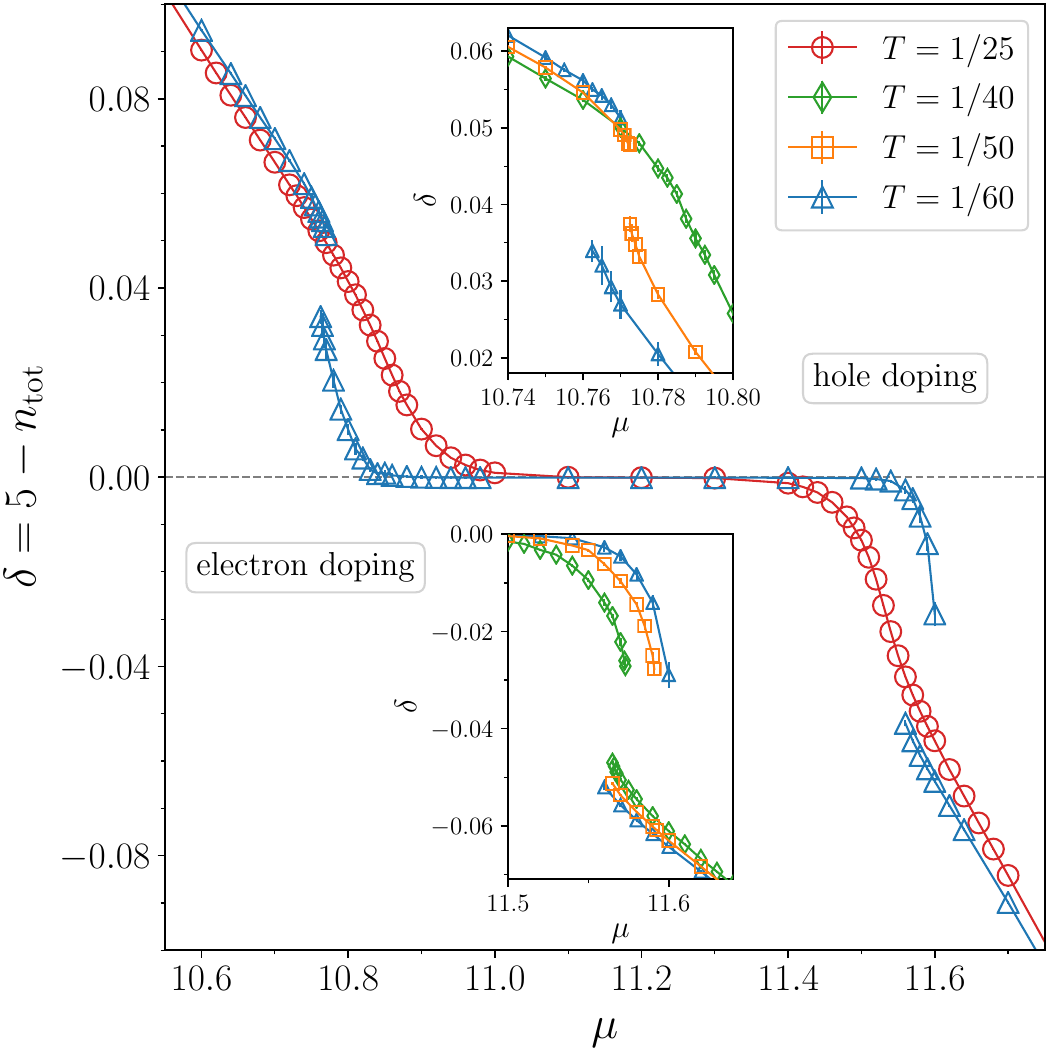}}
\caption{Doping $\delta= 5 -n_{\rm tot}$ vs chemical potential $\mu$ for different values of the temperature $T$. Hole doping corresponds to $\delta>0$ and electron doping corresponds to $\delta<0$. The plateau at $n_{\rm tot}=5$ indicates the incompressible charge-transfer insulator. At low temperatures, hysteresis loops develop at finite doping, for both hole and electron doping, as shown in the insets. Hysteresis marks the first-order transition between a strongly correlated pseudogap at small absolute dopings and a correlated metal at larger absolute dopings. Data are for the set of parameters of Eq.~\ref{eq:modelparameters} and $U_{d} =11.1$. 
}
\label{fig:nVSmu}
\end{figure}

Figure~\ref{fig:nVSmu} shows the doping $\delta$ as a function of the chemical potential $\mu$ for different values of temperature $T$. 
The doping is defined as $\delta=5-n_{\rm tot}$, therefore the system is electron-doped for negative $\delta$ and is hole-doped for positive $\delta$. 

At $n_{\rm tot}=5$, or equivalently $\delta=0$, the system is a charge-transfer insulator, as revealed by the plateau in the isotherms $\delta(\mu)$. The plateau in the $\delta(\mu)$ curves widens with decreasing temperature. The shape of the isotherms $\delta(\mu)$ is asymmetric upon adding or removing electrons from the charge-transfer insulator. 
 
Upon increasing or decreasing $\mu$, the isotherms $\delta(\mu)$ evolve from $\delta=0$ to a finite value of $\delta$. Hence the system becomes metallic since electrons are added or removed from the charge-transfer insulator. The metallic state obtained upon electron or hole doping the charge-transfer insulator has the features of a strongly correlated pseudogap, as we will discuss in Sec.~\ref{sec:dos}. The isotherms $\delta(\mu)$ evolve continuously from $\delta=0$ to a finite value of $\delta$, hence the transition between charge-transfer insulator and pseudogap is continuous. 

Upon further increasing or decreasing $\mu$, the doping further raises in absolute value, and the system evolves from a  pseudogap state to a more conventional correlated metallic state. At low temperatures ($T=1/50, 1/60$, orange and blue curves, respectively), the $\delta(\mu)$ curves show hysteresis loops as a function of $\mu$. Hysteresis loops appear when sweeping up and down the chemical potential and signals a first-order transition. The branches of the hysteresis loops show jumps at finite doping, indicating that the first-order transition occurs at finite doping, for both electron and hole doping. 
By increasing temperature, the hysteresis loops close and the $\delta(\mu)$ curves become continuous and show a sigmoidal shape, both for electron and hole doping ($T=1/25$, red curve). Hence, the overall behavior of the isotherms $\delta(\mu)$ implies that the pseudogap to correlated metal first-order transition terminates in a critical endpoint at finite temperature and finite doping. At the critical point, the $\delta(\mu)$ curve eventually displays a vertical tangent. 

To summarise, upon either electron or hole doping, at low temperature the system undergoes a purely electronic two-stage doping-driven metal insulator transition: a continuous charge-transfer insulator to pseudogap transition at $\delta=0$, and a discontinuous pseudpgap to correlated metal transition at finite $\delta$. 
This two-stage doping-driven metal insulator transition also occurs in the doped Mott insulator realised by the single-band 2D Hubbard model solved with CDMFT~\cite{sht, sht2}, as pointed out in Ref.~\cite{Lorenzo3band}.

\subsection{Peak in entropy and enhanced charge compressibility}

The behavior of the isotherms $\delta(\mu)$ encodes information about the thermodynamic entropy and the charge compressibility. 
First, the isotherms $\delta(\mu)$ cross at finite doping, both for electron and hole doping. The change of sign of $(d\delta/dT)_{\mu}$ reflects a vanishing expansion coefficient $(dn_{\rm tot}/dT)_{\mu}$ and thus, by the Maxwell relation $(dn_{\rm tot}/dT)_{\mu}=(ds/d\mu)_T$, an extremum (a maximum) in the thermodynamic entropy $s$ as a function of $\mu$ and hence doping. This feature also occurs in the single-band 2D Hubbard model~\cite{sht, sht2, Caitlin:PRXQ2020, CaitlinPNAS2021, MikelsonThermodynamics:2009} and is compatible with thermodynamic entropy measurements in the normal state of hole-doped cuprates~\cite{LoramJPCS2001, Tallon:2004}.

\begin{figure}[t!]
\centering{
\includegraphics[width=1.\linewidth]{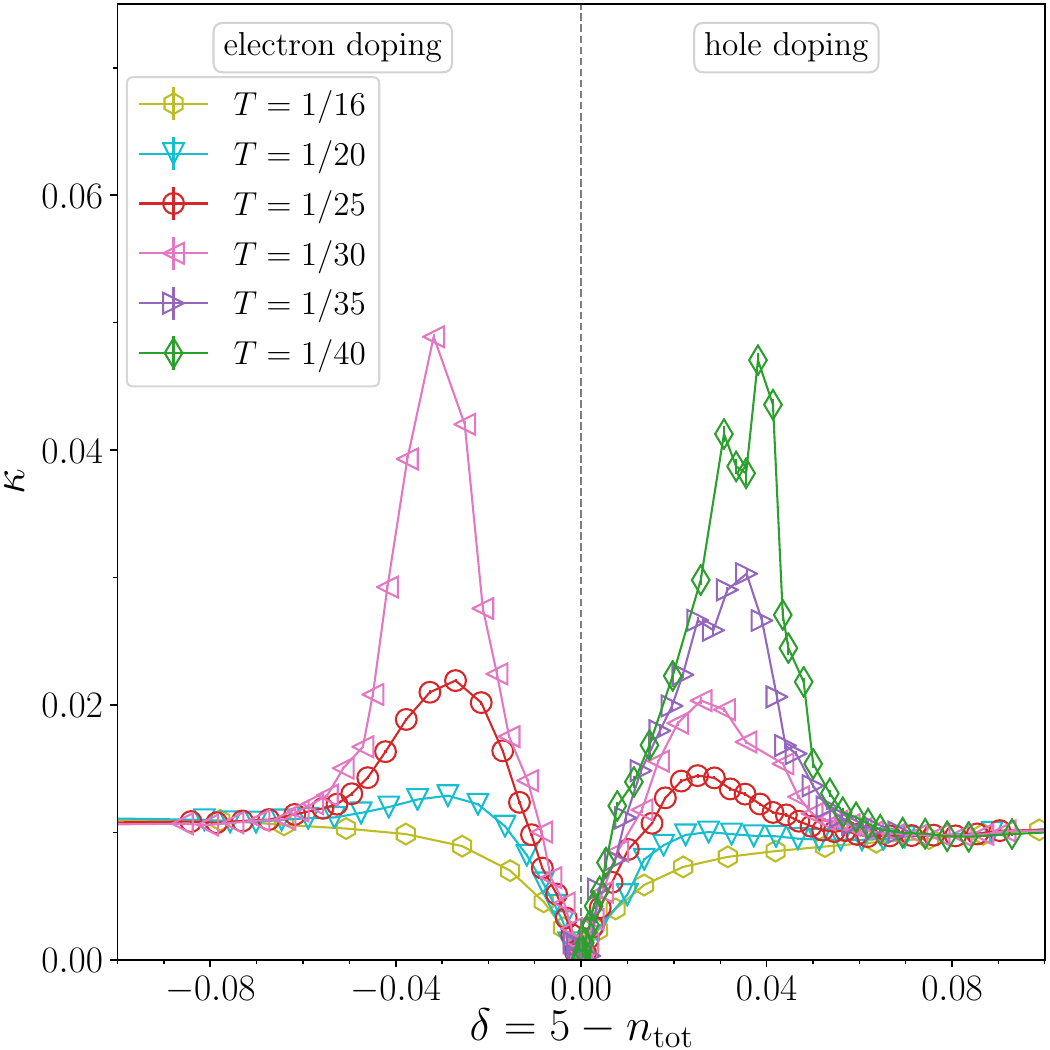}}
\caption{Isothermal charge compressibility $\kappa = n_{\rm tot}^{-2} (dn_{\rm tot}/d\mu)_T$ vs doping $\delta$ for different values of the temperature $T$. Electron doping is for $\delta<0$, hole doping is for $\delta>0$. Upon either electron or hole doping, $\kappa(\delta)$ develops a temperature dependent peak. Data are for the set of parameters of Eq.~\ref{eq:modelparameters} and $U_{d} =11.1$. 
}
\label{fig:compressibility}
\end{figure}

Second, the derivative of the $\delta(\mu)$ curves is related to the charge compressibility $\kappa = n_{\rm tot}^{-2} (dn_{\rm tot}/d\mu)_T$. Figure~\ref{fig:compressibility} shows the charge compressibility as a function of doping for different values of the temperature. At zero doping, $\kappa=0$, implying that the system is an incompressible charge-transfer insulator. Upon adding or removing electrons from the charge-transfer insulator, the charge compressibility increases with increasing doping. Upon reducing the temperature, the charge compressibility develops a peak at finite doping, for both electron and hope doping. The shape of the peak is asymmetric upon hole or electron doping. The peak in $\kappa(\delta)$ increases in magnitude with decreasing the temperature and eventually it diverges at the critical endpoint of the pseudogap to metal first-order transition, where $(dn_{\rm tot}/d\mu)$ tends to infinity. Physically, the increased charge compressibility reflects the amplification of the thermodynamic density fluctuations emanating from the critical endpoint -- the analogue of the critical opalescence, as discussed in Ref.~\cite{CaitlinOpalescence}. Intriguingly, Ref.~\cite{CampiOpalescence:2022} pointed out that the analogue of critical opalescence is consistent with the experimental observation of scale-invariant structural organization of oxygen interstitials at optimum doping found in the HgBa$_2$CuO$_{4+y}$ and La$_2$CuO$_{4+y}$ hole-doped cuprates~\cite{Bianconi:2010, CampiOpalescence:2022}.

\section{Temperature-doping phase diagram} 
\label{sec:phasediagram}

\begin{figure}[ht!]
\centering{
\includegraphics[width=1.\linewidth]{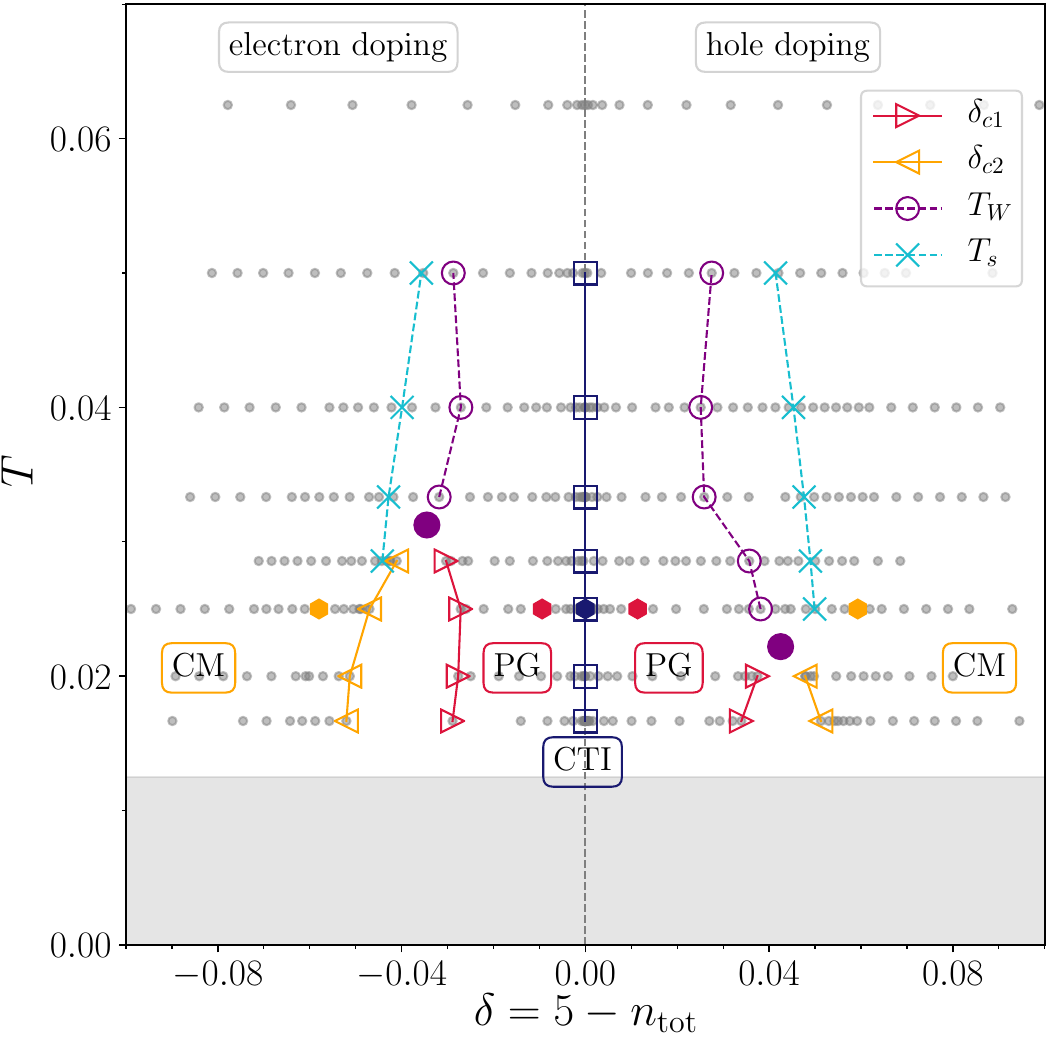}}
\caption{Temperature $T$ vs doping $\delta$ normal state phase diagram of the Emery model solved with CDMFT. Data are for the set of parameters of Eq.~\ref{eq:modelparameters} and $U_{d} =11.1$. Left: electron doping, right: hole doping. 
Dark blue squares show the incompressible charge-transfer insulator (CTI) at $\delta=0$. The spinodal line marking the pseudogap (PG) boundary is shown by open red triangles. The spinodal line marking the correlated metal (CM) boundary is shown by open orange triangles. The spinodal lines eventually merge at a critical end point shown by a solid purple circle. The Widom line, determined from the locus of maximum charge compressibility $\kappa$ vs $\delta$ in Fig.~\ref{fig:compressibility}, extends from the critical end point and is indicated by a dashed purple line with open circles. The cyan dashed line with crosses shows the locus of maximum thermodynamic entropy $s$ vs $\delta$, as calculated by the crossing of the isotherms $\delta(\mu)$. These features occur on both the electron and hole-doped side of the phase diagram but are asymmetrical. Gray filled circles indicate the datapoints studied in our work. 
Filled hexagons denote the datapoints discussed in Sec.~\ref{sec:localDOS}. 
Shaded grey area marks the inaccessible region to our calculations due to the Monte Carlo sign problem. 
}
\label{fig:phasediagramT}
\end{figure}

The behavior of the doping $\delta(\mu)$ and the resulting charge compressibility $\kappa(\delta)$ for different temperatures discussed in Sec.~\ref{sec:occupation} allows us to construct the temperature versus doping phase diagram shown in Fig.~\ref{fig:phasediagramT}. The gray filled circles indicate the datapoints explored in our study. If at zero doping, $\kappa$ is zero, the system is an incompressible charge-transfer insulator (dark blue squares). For either electron or hole doping, the isotherms $\delta(\mu)$ of Fig.~\ref{fig:nVSmu} evolves continuously from $\delta=0$ to a finite value of $\delta$, hence the phase transition at $\delta=0$ between the charge-transfer insulator and the pseudogap is continuous. 

At finite dopings and low temperatures, the $\delta(\mu)$ isotherms of Fig.~\ref{fig:nVSmu} show hysteresis loops, which signal a first-order transition, for either hole or electron doping. The discontinuous jumps in $\delta(\mu)$ denote the disappearance of the strongly correlated pseudogap state at $\delta_{c1}$ and of the correlated metallic state at $\delta_{c2}$. 
By tracking the position of $\delta_{c1}$ and $\delta_{c2}$ for each temperature, one obtains the spinodal lines with red and orange open triangles in Fig.~\ref{fig:phasediagramT}. 

The spinodal lines merge at a second order critical endpoint, for either hole or electron doping. The two filled purple circles are our estimates for the critical end points on the hole and electron doped side of the phase diagram. As a result, the pseudogap to metal transition is a purely electronic transition. Its critical endpoint should belong to the Ising universality class, as shown for the metal to Mott insulator transition~\cite{Castellani:PRL1979, gabiEPJB, lange, Nagaosa:JPSJ2003, patrickCritical}. 

At temperatures above the critical endpoint of the pseudogap to metal transition there is only one phase. However, quite generally from a critical endpoint emerges a supercritical crossover defined by the locus of maximum correlation length, known as Widom line~\cite{water1, supercritical}, which produces extrema in thermodynamic response functions~\cite{water1}. In practice, the Widom line is often approximated by the locus of extrema of some thermodynamic response functions~\cite{water1, Luo_PRL2014, Giovanni:PRD2024} -- in Fig.~\ref{fig:phasediagramT}, we approximate the Widom line by the locus of maximum charge compressibility (dashed purple line with open circles).   
A peak in the charge compressibility is only one example of thermodynamic anomalies as a function of doping emerging from the pseudogap to metal critical endpoint~\cite{ssht}. In the context of the single-band 2D Hubbard model, anomalies in the form of extrema as a function of doping have been found in the electronic specific heat~\cite{Giovanni:PRBcv}, density fluctuations~\cite{CaitlinOpalescence}, sound velocity~\cite{CaitlinSoundVelocity}, and entanglement properties~\cite{Caitlin:PRXQ2020}. 
These anomalies can be detected in experiments. As already mentioned in Sec.~\ref{sec:occupation}, anomalies in the density fluctuations have been observed at optimum doping in HgBa$_2$CuO$_{4+y}$ and La$_2$CuO$_{4+y}$ hole-doped cuprates~\cite{Bianconi:2010, CampiOpalescence:2022}. A pronounced peak in the low temperature electronic specific heat has been measured at the doping where the pseudogap ends in the normal state of hole-doped cuprates~\cite{Michon:Cv2018}. 

Next, let us compare the hole and electron doping sides of the $T-\delta$ phase diagram. They appear qualitatively similar, but quantitatively different. In both cases, the system shows a first order transition between a pseudogap and a correlated metal which terminates in a critical end point at finite doping and finite temperature, and from which emerges a supercritical crossover characterised by enhanced density fluctuations. By changing the interaction strength $U_{d}$, in Section~\ref{sec:Ud} we will show that the first order transition at hole doping and at electron doping are connected in the $U_d-T-\delta$ phase diagram, thereby revealing their common origin in the Mott physics in two dimensions. In other words, in the $U_{d}-T-\delta$ diagram, there is a surface of first-order transitions.  

For either electron or hole doping, the pseudogap to metal transition is followed at large doping by a maximum of the thermodynamic entropy $s(\delta)$. The locus of the maximum in $s(\delta)$ at different temperatures, as determined by the zero of the expansion coefficient $(dn_{\rm tot}/dT)_{\mu}$, is shown by a cyan dotted line with crosses: it is only weakly doping dependent and extends up to high temperatures. Physically, the peak in the thermodynamic entropy arises from the localization-delocalization physics of doped Mott insulators, as discussed in Ref.~\cite{MikelsonThermodynamics:2009, sht, sht2, Alexis:2019, Caitlin:PRXQ2020, Lenihan:PRL2021} in the context of the single-band 2D Hubbard model. 
These key features of the thermodynamic entropy $s$ -- a doping-independent peak in $s(\delta)$ close to the value of doping where the pseudogap ends -- are compatible with experimental results in hole-doped cuprates~\cite{LoramJPCS2001, Tallon:2004}. 

Let us now turn to the quantitative differences between the two sides of the $T-\delta$ phase diagram. The temperature and doping position of the critical end point is asymmetric between electron and hole doping. The temperature of the critical end point is higher on the electron-doped side of the phase diagram. The origin of this asymmetry is not evident to us. 
A similar electron-hole asymmetry in the pseudogap to metal critical endpoint occurs in the single-band 2D Hubbard model on a triangular lattice~\cite{Downey:PRB2024}. This suggests that this asymmetry may be linked to the shape of the Fermi surface upon doping. However further investigations are needed to clarify this aspect. A possible hint may come from the study of the electron-hole asymmetry in the pseudogap to metal critical endpoint in the 2D Hubbard model with next-nearest neighbor hopping on a square lattice~\cite{LorenzoSC}. 

To summarise, upon either electron or hole doping, there is a purely electronic transition between a strongly correlated pseudogap and a more conventional correlated metal. This transition is first order and ends in a critical endpoint, from which a supercritical crossover (the Widom line) emerges. The transition is followed at large doping in absolute value by a maximum in the thermodynamic entropy vs doping. 
The main features of the $T-\delta$ phase diagram of Fig.~\ref{fig:phasediagramT} confirm those found in the prior work of Ref.~\cite{Lorenzo3band} for hole doping and extend them to electron doping. 

All these features also occur in the doped Mott insulator realised by the single-band 2D Hubbard model solved with CDMFT~\cite{sht, sht2, ssht, Caitlin:PRXQ2020, CaitlinPNAS2021}. As pointed out in Ref.~\cite{Lorenzo3band}, this thermodynamic landscape is robust against microscopic bandstructure parameters and thus is an emergent feature of a doped correlated insulator in two dimensions - a doped Mott insulator realised by the 2D Hubbard model and a doped charge-transfer insulator realised by the Emery model. 

Let us also stress that the similarities found between the $T-\delta$ phase diagram of the {\it electron-doped} charge-transfer insulator realised by the Emery model and the doped Mott-Hubbard insulator realised by the 2D Hubbard model are expected on physical ground: upon introducing electrons into the charge-transfer insulator, the charge-transfer band just below the Fermi energy is almost full of electrons, and thus almost electrically and magnetically inert, serving mainly as a source of electrons for the delocalisation of the electrons among the Cu $3d$ orbitals. 

On the other hand, the similarities found between the $T-\delta$ phase diagram of the {\it hole-doped} charge-transfer insulator realised by the Emery model and the doped Mott-Hubbard insulator realised by the 2D Hubbard model are arguably less expected: upon adding holes into the charge-transfer insulator, the holes have a strong mixed $d$-$p$ character. However, in this regime, Zhang and Rice~\cite{ZhangRice1988} have long ago argued that singlets develop between copper and oxygen orbitals -- the so-called Zhang-Rice singlets, i.e. singlet states formed by an intrinsic hole in the Cu $3d$ orbital and a doped hole on the surrounding oxygen $2p$ orbitals. These singlets delocalise throughout the lattice formed by the Cu sites and thus their physics is well described by a single-band Hubbard model~\cite{ZhangRice1988}.

\section{Phase characterization: local density of states and spectral function at the Fermi level}
\label{sec:dos}

\begin{figure}
\centering{
\includegraphics[width=1.\linewidth]{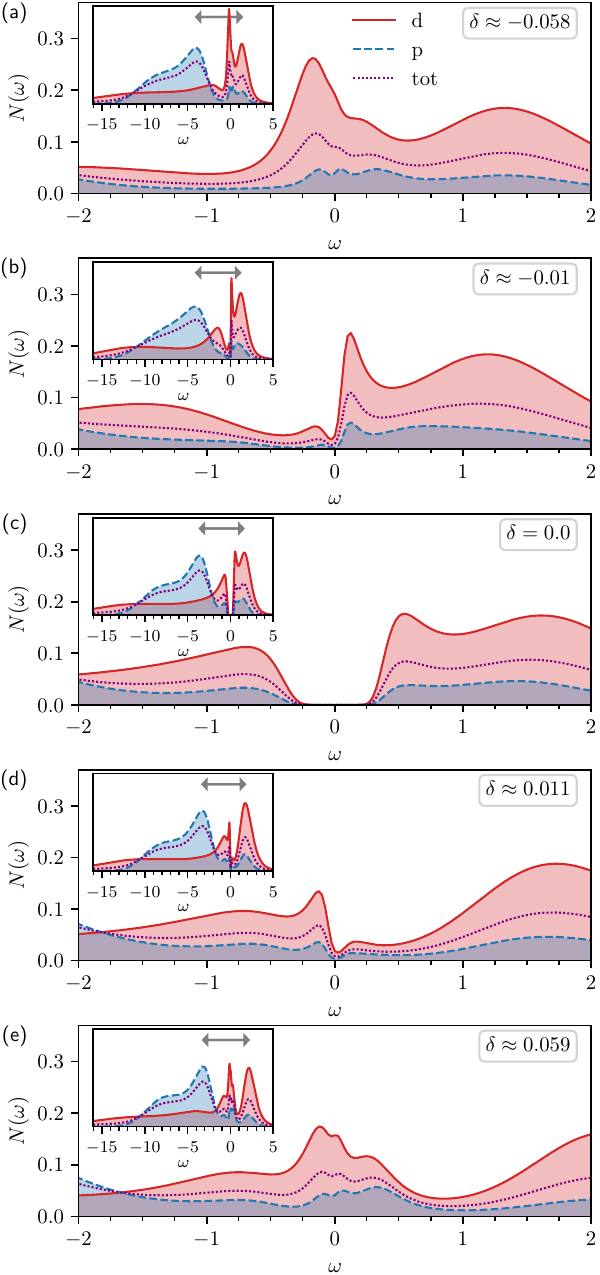}}
\caption{Projected local density of states $N(\omega)$ on the Cu $3d$ orbital (red solid line) and on the degenerate O $2p_x, 2p_y$ orbitals (blue dashed line), as well as the total density of states $N_{\rm tot}(\omega)=(N_d(\omega) +2N_p(\omega))/3$ (purple dotted line), for several doping levels, corresponding to the hexagons in Figure~\ref{fig:phasediagramT}. Each DOS is normalised to 1. 
(a),(b): electron doping; (c) charge-transfer insulator; (d),(e): hole doping. Data are for the set of parameters of Eq.~\ref{eq:modelparameters}, $U_{d} =11.1$ and $T=1/40$. The grey double arrow in the insets marks the distance $\Delta_g$ between the charge-transfer band and the upper Hubbard band on $N_{\rm tot}(\omega)$. The doping evolution of $\Delta_g$ is shown in Fig.~\ref{fig:dosPG}(d). 
}
\label{fig:dos}
\end{figure}
\begin{figure}
\centering{
\includegraphics[width=1.\linewidth]{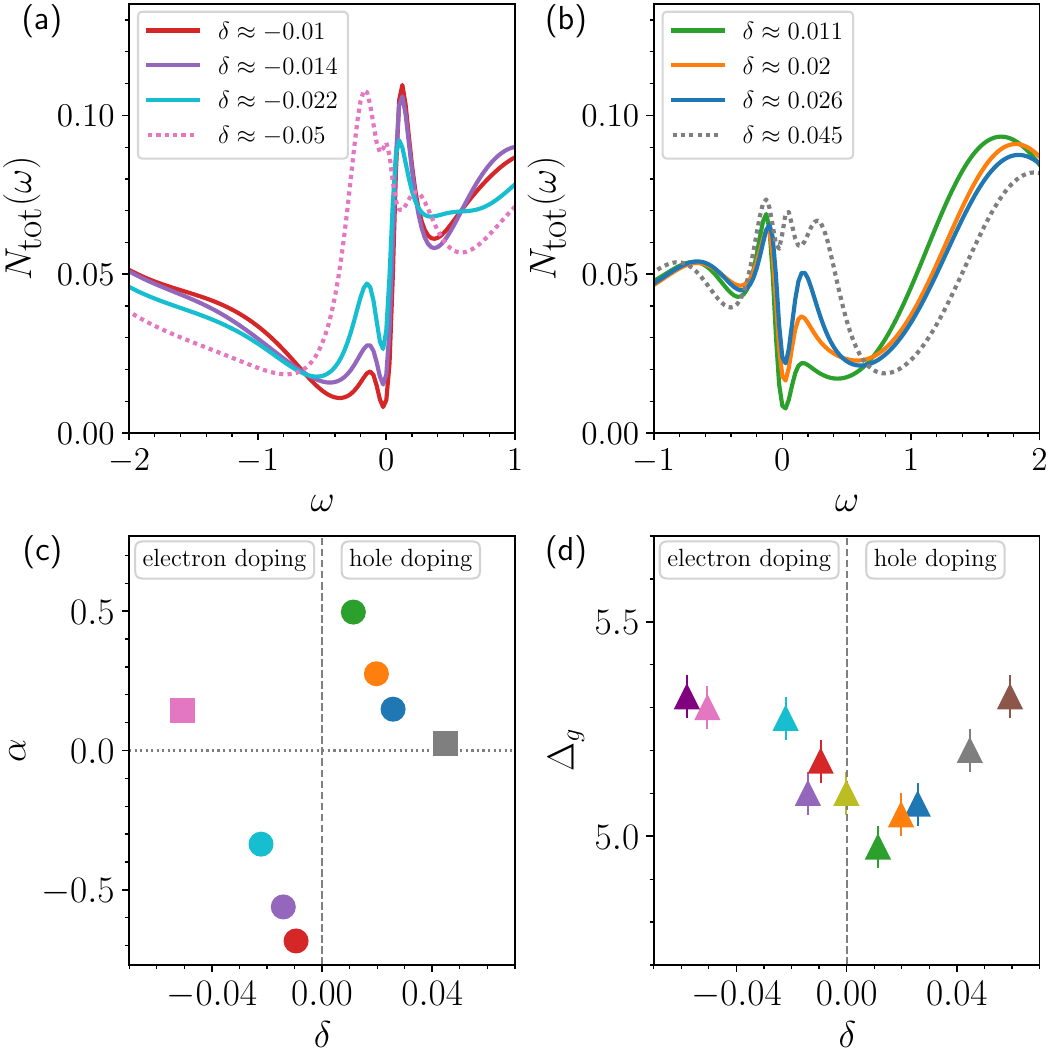}}
\caption{(a), (b) $N_{\rm tot}(\omega)=(N_d(\omega) +2N_p(\omega))/3$ for several electron and hole doping values [panels (a) and (b), respectively]. Solid lines show $N_{\rm tot}(\omega)$ in the pseudogap state. For comparison, the dotted lines show $N_{\rm tot}(\omega)$ in the correlated metal state. (c) Low frequency asymmetry parameter $\alpha$ (see definition in the text) vs doping. It measures the low frequency pseudogap electron-hole asymmetry. $\alpha$ changes sign when moving from electron to hole doping. (d) Distance $\Delta_g$ between the charge-transfer band and the upper Hubbard band on $N_{\rm tot}(\omega)$ (see grey double arrows in Fig.~\ref{fig:dos}) as a function of doping.
Data are for the set of parameters of Eq.~\ref{eq:modelparameters}, $U_{d} =11.1$ and $T=1/40$.
}
\label{fig:dosPG}
\end{figure}

The analysis in Sections~\ref{sec:occupation} and \ref{sec:phasediagram} reveals the existence of a first-order transition between two compressible metals at finite electron and hole doping. To further characterise the two metallic states separated by the transition and to show that they have the features of a strongly correlated pseudogap and of a correlated metal, we can turn to the study of the local density of states (DOS) and of the spectral function at the Fermi level. The strategy of analyzing the local DOS has already been adopted in Ref.~\cite{Lorenzo3band} for the hole-doped Emery model. Here we extend it to the electron doped case. 

\subsection{Local density of states}
\label{sec:localDOS}

Figure~\ref{fig:dos} shows the projected local density of states $N(\omega)=-\frac{1}{\pi} \textrm{Im}G(\omega)$ on the copper $3d$ orbital (red solid line) and on the oxygen $2p_x, 2p_y$ orbitals (blue dashed line), as well as the total density of states $N_{\rm tot}(\omega)=(N_d(\omega) +2N_p(\omega))/3$ (purple dotted line) at the low temperature $T=1/40$ and for different values of doping, corresponding to the filled hexagons in the phase diagram of Fig.~\ref{fig:phasediagramT}. The main panels contain the low frequency part of $N(\omega)$, whereas the inset of each panel shows the full frequency range of $N(\omega)$. The analytical continuation from Matsubara to real frequencies has been performed using the method of Ref.~\cite{DominicMEM}. 

At zero doping [panel (c)] the correlation strength $U_d$ opens up a charge-transfer insulating gap between the charge-transfer band and the upper Hubbard band. The bands separated by the gap have a large mixed $d$-$p$ character. 

At a doping larger in absolute value than the first order transition [panels (a) and (e)], $N(\omega)$ shows a narrow peak close to the Fermi energy, indicating a correlated metallic state. The peak has predominant $d$ character at electron doping, whereas it has large mixed $d$-$p$ character at hole doping. 

At small either electron or hole doping [panels (b) and (d)], the electronic correlations on the Cu $3d$ orbital produce a substantial redistribution of spectral weight from high to low frequencies. 
At low frequencies, $N(\omega)$ develops a pseudogap, i.e. a partial depression in the density of states, between a peak below and a peak above the Fermi energy.  The pseudogap has mixed $d$-$p$ character, which is larger at hole doping than at electron doping. The pseudogap shows a two-peak structure with a marked asymmetric shape as a function of frequency $\omega$, which is greater for the $d$-DOS: for hole doping there is a more pronounced peak below the Fermi energy, whereas for electron doping, there is a more pronounced peak above the Fermi energy. Therefore, upon changing the sign of the dopant carriers, the particle-hole asymmetry of the pseudogap, i.e. the height difference of the peaks forming the pseudogap, is reversed. 

To further analyse the low frequency electron-hole asymmetry of the pseudogap, Fig.~\ref{fig:dosPG}(a,b) shows the total DOS for several values of electron and hole doping (panel (a) and (b), respectively). First, upon increasing the value of absolute doping, the pseudogap fills in. Second, in Fig.~\ref{fig:dosPG}(c) we quantify the pseudogap low frequency electron-hole asymmetry by introducing the asymmetry parameter~\cite{Zhong:PRL2020} $\alpha= (A_{-}-A_{+})/(A_{+}+A_{-})$, where $A_{\pm}$ are the occupied and unoccupied areas of $N_{\rm tot}(\omega)$ within a small interval $[ 0,\omega_\pm ]= \pm0.25$. Upon changing between electron and hole doping, the asymmetry parameter $\alpha$ changes sign. Upon increasing the value of absolute doping, $|\alpha|$ is reduced. 
Note that the transfer of spectral weight that accompanies the development of the pseudogap is not appreciably altering the high energy features of $N(\omega)$. To show this, Fig.~\ref{fig:dosPG}(d) displays the doping evolution of the distance $\Delta_g$ between the charge-transfer band and the upper Hubbard band (see gray double arrow in insets of Fig.~\ref{fig:dos}). Upon doping, $\Delta_g$ has a variation of $\approx 0.5$ which is small compared to the overall bandwidth of $N_{\rm tot}(\omega)$, which is $\approx 20$ (see insets of Fig.~\ref{fig:dos}). 
In Section~\ref{sec:SLCO} we shall show that all these findings are in qualitative agreement with the experimental results on ambipolar SLCO cuprate films in Ref.~\cite{Zhong:PRL2020}.

\subsection{Spectral function at the Fermi level}

\begin{figure*}[ht]
\centering{
\includegraphics[width=1.\linewidth]{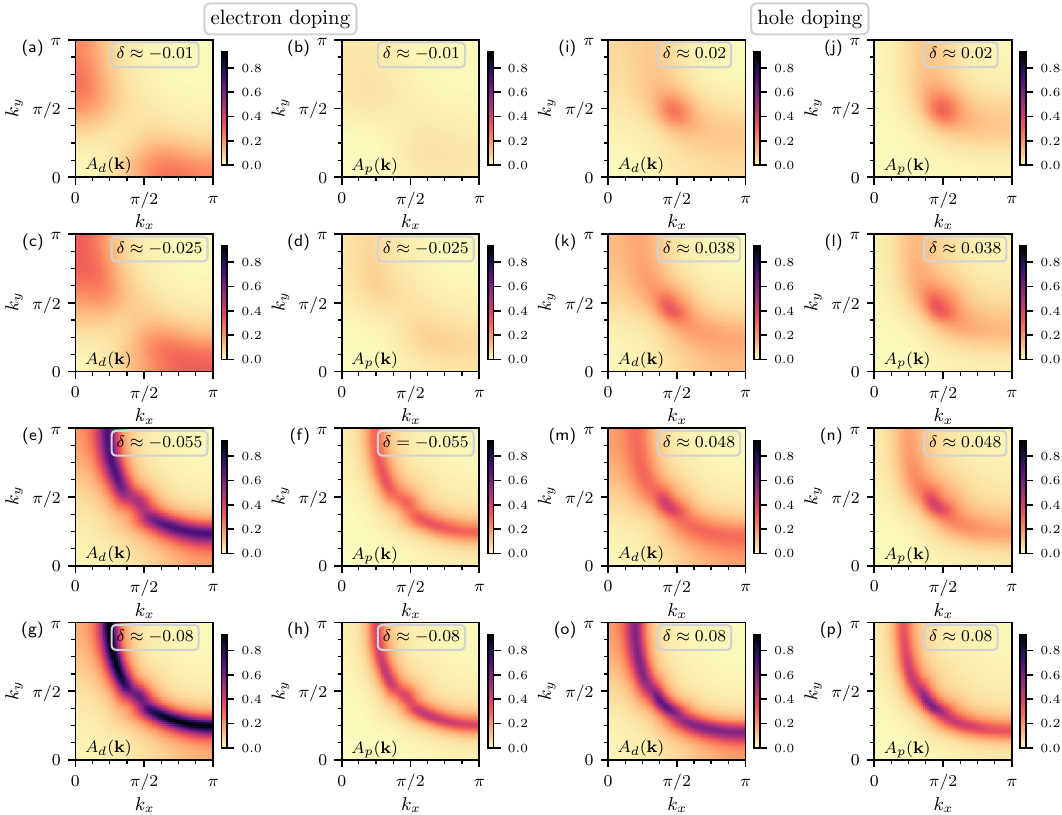}
}
\caption{Intensity plot of the momentum resolved spectral function at the first Matsubara frequency of the Cu $3d$ orbital ($A_d({\bf k}$)) and of the oxygen $2p_x, 2p_y$ orbitals ($A_p({\bf k})$) in the first quadrant of the Brillouin zone. Data are for the set of parameters of Eq.~\ref{eq:modelparameters}, $U_{d} =11.1$ and $T=1/50$, which lies below the temperature of the pseudogap to metal critical endpoint (see Fig.~\ref{fig:phasediagramT}). Panels (a)-(h) correspond to different values of electron doping. Panels (i)-(p) correspond to different values of hole doping. 
}
\label{fig:Akw1}
\end{figure*}

Figure~\ref{fig:Akw1} shows the intensity plot of the spectral function at the first Matsubara frequency $\omega_0=i\pi T$ of the Cu $3d$ orbital ($A_d({\bf k})=-\frac{1}{\pi}\textrm{Im}G_d({\bf k}, i\omega_0)$) and of the O $2p_x, 2p_y$ orbitals ($A_p({\bf k})=-\frac{1}{\pi}\textrm{Im}G_p({\bf k}, i\omega_0)$) at the low temperature $T=1/50$ and for different values of doping. At low temperature, this quantity is a proxy of the spectral function at the Fermi level. 
Data are obtained for the set of parameters of Eq.~\ref{eq:modelparameters} and $U_{d} =11.1$. 
To obtain the translation invariant lattice Green's functions $G_d({\bf k}, i\omega_0)$ and $G_p({\bf k}, i\omega_0)$ from the cluster self-energy $\boldsymbol{\Sigma}_{\rm cl} (i\omega_0)$, we use the Green's function periodization scheme~\cite{Senechal:PRB2002}. 

Let us first discuss the correlated metal at values of doping larger in absolute value than the first-order transition [third and fourth rows, i.e. panels (e)-(h) for electron doping and panels (m)-(p) for hole doping]. 
At large $|\delta|$ (fourth row), upon either electron or hole doping, the intensity along the locus of maximum spectral function is large, indicating long-lived excitations, and approximately uniform in ${\bf k}$. 
Upon electron doping, the overall intensity of $A_d({\bf k})$ is larger than that of $A_p({\bf k})$. On the other hand, upon hole doping, the overall intensity of $A_d({\bf k})$ is comparable to that of $A_p({\bf k})$. This reflects the asymmetry regarding which orbital the dopant carriers enter. 
Upon reducing doping in absolute value [third row], the overall spectral intensity decreases and broadens as a results of the increased correlations. 

Next, let us discuss the pseudogap state at small values of either electron or hole doping [first and second rows, i.e. panels (a)-(d) for electron doping and panels (i)-(l) for hole doping]. Upon either electron or hole doping, the intensity along the locus of maximum spectral function is overall small and strongly ${\bf k}$ dependent. This is the hallmark signature of the strongly correlated pseudogap state on the spectral function within cluster DMFT methods, as shown by intensive studies on the single band 2D Hubbard model on a square lattice~\cite{st, civelliBreakup, kyung, Macridin:2006, sakaiPRL, michelPRB, michelCFR} (for a recent review, see Ref.~\cite{Sakai:JPSJ2023}). Our results extend this finding to the Emery model, and show that a strong ${\bf k}$ dependent depletion of states at the Fermi level occurs at either electron or hole doping. 

There is however a quantitative asymmetry between electron and hole doping regarding the region in ${\bf k}$ where the depletion of spectral intensity takes place. For hole doping, the spectral intensity is concentrated along the zone diagonal (i.e., the so-called nodal direction) and is gradually reduced upon approaching the zone boundary (i.e., the so-called antinodal direction). In sharp contrast, for electron doping the spectral intensity is concentrated along the zone boundary and is progressively depleted upon approaching the zone diagonal. 

To further characterize the first-order transition between strongly correlated pseudogap and correlated metal, the second and third rows of Fig.~\ref{fig:Akw1} show the spectral intensity in the coexistence region: panels (c)-(d) [(k)-(l)] and (e)-(f) [(m)-(n)] have the same chemical potential $\mu$, but different doping levels (see Fig.~\ref{fig:nVSmu}) and different spectral functions. Across the pseudogap to metal first-order transition there is a dramatic depletion and broadening of spectral intensity. The depletion of spectral intensity is especially strong close to the zone diagonal for electron doping, and away from the zone diagonal for hole doping.  

Finally, we caution that the calculation of the momentum resolved spectral function in Fig.~\ref{fig:Akw1} relies on the periodization -- a procedure applied to the converged CDMFT results to extract the translation invariant lattice Green's function from the cluster self-energy. The periodization is an ill-defined procedure and different periodization schemes have been proposed. Here we used the Green's function periodization scheme~\cite{Senechal:PRB2002}. It has long been debated whether the spectral intensity features are an artefact of the periodization or a result of CDMFT. Recently, Ref.~\cite{Verret:PRB2022} has demonstrated that CDMFT cannot resolve the question of Fermi arcs vs hole pockets, regardless of the periodization used. Nonetheless, the key features of the spectral intensity on the single band 2D Hubbard model solved with CDMFT are compatible with those revealed by other numerical methods~\cite{Michel:Science2024}. Note also that the phase characterization by the local DOS in Sec.~\ref{sec:localDOS} does not rely on the periodization procedure.

\section{Redistribution of the orbital character of the dopants}
\label{sec:holecontent}
\begin{figure}[ht]
\centering{
\includegraphics[width=1.\linewidth]{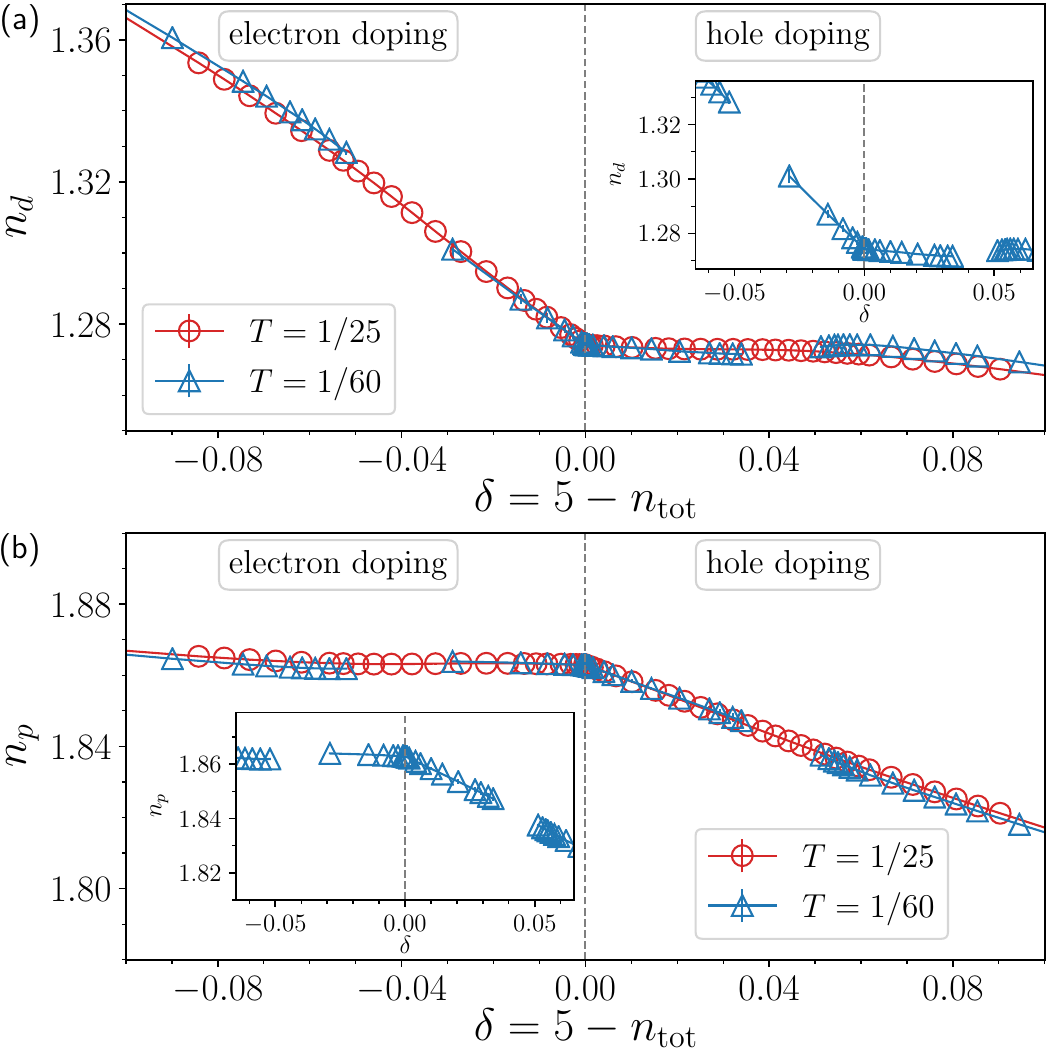}}
\caption{Partial occupation of the copper $3d$ orbital, $n_d$ [panel (a)], and of the equivalent oxygen $2p_x, 2p_y$ orbitals, $n_p$ [panel (b)], vs doping $\delta$ for two values of temperature above and below the pseudogap-metal critical endpoint. Left: electron doping, right: hole doping. Inset in each panel zooms into the pseudogap to metal transition region, for $T=1/60$. Data are for the set of parameters of Eq.~\ref{eq:modelparameters} and $U_{d} =11.1$.
}
\label{fig:asymmetryOcc}
\end{figure}

The preceding sections~\ref{sec:occupation}-\ref{sec:dos} showed that there is no qualitative asymmetry in the nature of the metal to charge-transfer insulator transition induced by either hole or electron doping. This section focuses on the second main question of our work, namely how the orbital character of the dopant carriers evolves with either doping path. 

Figure~\ref{fig:asymmetryOcc} shows the partial occupation of the copper $3d$ orbital, $n_d$, and of the oxygen $2p_x, 2p_y$ orbitals (which are equivalent), $n_p$, as a function of doping for different values of temperature. Note that in our convention the partial occupation ranges between 0 and 2, where 0 means that the orbital is empty, and 2 means that the orbital is fully occupied with two electrons. 

First, let us consider zero doping. In the absence of the hybridization, $t_{pd}=0$, the interaction strength $U_d$ splits the copper $3d$ level into two levels at $\epsilon_d$ and $\epsilon_d+U_d$. In a charge-transfer insulator the copper level $\epsilon_d +U_d$ lies above the oxygen $2p$ band, hence the copper occupation $n_d$ is equal to $1$ and the oxygen occupation $n_p$ is equal to $2$. Hence the total occupation is equal to $n_{\rm tot} =n_d + 2n_p=5$. 
Due to the finite hybridization, $t_{pd}\neq 0$, the charge-transfer insulator is then realised for $n_d = 1 +2\eta$ and correspondingly $n_p=2-\eta$ (for the set of parameters in Fig.~\ref{fig:asymmetryOcc} one finds $n_d \approx 1.274$ and $n_p \approx 1.863$). In other words, the magnitude of the partial occupation is slightly more than half filling for $n_d$ and is almost fully filled for $n_p$, as schematically represented in the central panel of Fig.~\ref{fig:sketch}(b).  

Let us now turn to the doped case. Our results show that there is an asymmetry regarding which orbital the dopant carriers go to. 
Upon electron doping [left part of Fig.~\ref{fig:asymmetryOcc}(a),(b)], since almost all the oxygen $2p$ orbitals are occupied, the added electrons mainly go to the copper $3d$ orbitals. Therefore $n_d$ increases with increasing in absolute value the electron doping, whereas the occupation of the oxygen $2p$ orbitals do not vary much with electron doping (see also the sketch on the left panel of Fig.~\ref{fig:sketch}(b)). 
On the contrary, upon hole doping [right part of Fig.~\ref{fig:asymmetryOcc}(a),(b)], the doped holes mainly go to occupy the oxygen $2p$ orbitals: the occupation of the copper $3d$ orbital remains basically unchanged upon hole doping, whereas the occupation of the oxygen $2p$ orbitals decreases with increasing hole doping (see also the sketch on the right panel of Fig.~\ref{fig:sketch}(b)). 

In Sec.~\ref{sec:occupation} we showed that at low temperature, upon either electron or hole doping, the system undergoes a pseudogap to metal first order transition at finite doping, revealed by discontinuous jumps in the total occupation $n_{\rm tot}(\mu)$. This first order transition is also visible in the partial occupancies $n_d$ and $n_p$ [blue triangles at $T=1/60$ in Fig.~\ref{fig:asymmetryOcc}(a),(b)]. Upon electron doping, the size of the jump of the partial occupancy is larger on the copper $3d$ orbitals, whereas upon hole doping the size of the jump of the partial occupancy is larger on the oxygen $2p$ orbitals.

\begin{figure}[t!]
\centering{
\includegraphics[width=1.\linewidth]{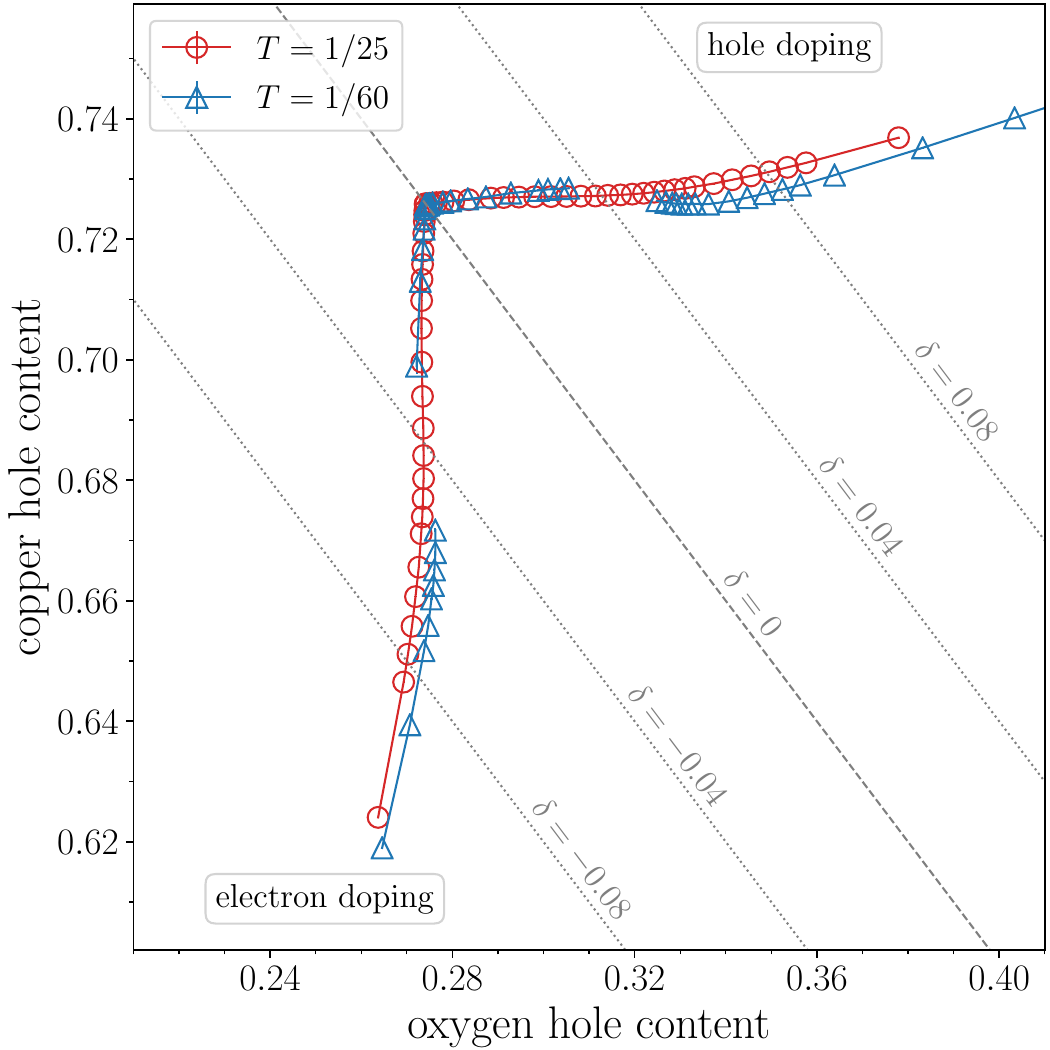}}
\caption{Copper hole content (defined as $2-n_d$ vs oxygen hole content (defined as $2(2-n_p)$, for two values of the temperature. The gray dashed line marks the charge-transfer insulator at $\delta=0$ and separates electron doping from hole doping (bottom left and top right of the graph, respectively). The gray dotted lines indicate constant electron or hole dopings. Data are for the set of parameters of Eq.~\ref{eq:modelparameters} and $U_{d} =11.1$.
}
\label{fig:holecontent}
\end{figure}

The data from both panels of Fig.~\ref{fig:asymmetryOcc} can be condensed into one figure. This is because $n_d$ and $n_p$ are related by $n_d +2n_p = n_{\rm tot} = 5 - \delta$. One can therefore plot $n_d$ vs $n_p$, which directly quantifies the redistribution of the orbital $d$-$p$ character of the dopants. Refs.~\cite{Jurkutat:PRB2014, Rybicki:NatComm2016, Nicolas:PNAS2021, Nicolas:Master, Davis:PNAS2022, Jurkutat:PNAS2023} have shown that insights can be gained from this analysis. 
To make contact with the experimental Refs.~\cite{Jurkutat:PRB2014, Rybicki:NatComm2016, Jurkutat:PNAS2023}, we consider the hole representation, i.e. rather than plotting $n_d$ vs $n_p$, in Fig.~\ref{fig:holecontent} we plot the copper hole content (defined as $2-n_d$) vs oxygen hole content (defined as $2(2-n_p)$), for two values of temperature. 
For example, the charge-transfer insulator is realised with five electrons, or equivalently one hole. The gray dashed line in Fig.~\ref{fig:holecontent} indicates the charge-transfer insulator at $\delta=0$ and separates electron doping (bottom left region of the graph) from hole doping (top right region of the graph). The gray dotted lines parallel to the $\delta=0$ line mark constant electron or hole dopings. Doping increases in absolute value by moving away from the $\delta=0$ line. 

At zero doping, the hole has a large mixed $d$-$p$ character, with a ratio of copper/oxygen hole content of approximately $0.726/0.274$. Physically, this means that the hole in the charge-transfer insulator is shared among the copper $3d$ and oxygen $2p$ orbitals, with a predominant weight on the Cu orbitals. 

Doping redistributes the hole content dramatically. This redistribution is strongly electron-hole asymmetric. Electron doping mainly decreases the copper hole content without changing much the oxygen hole content, especially not too far away from the first-order pseudogap to metal transition. Physically, this means that doped electrons mainly occupy the copper $3d$ orbitals. On the contrary, hole doping predominantly increases the oxygen hole content without changing much the copper hole content. Physically, this means that doped holes mainly enter the oxygen $2p$ orbitals.

\begin{figure}[t!]
\centering{
\includegraphics[width=1.\linewidth]{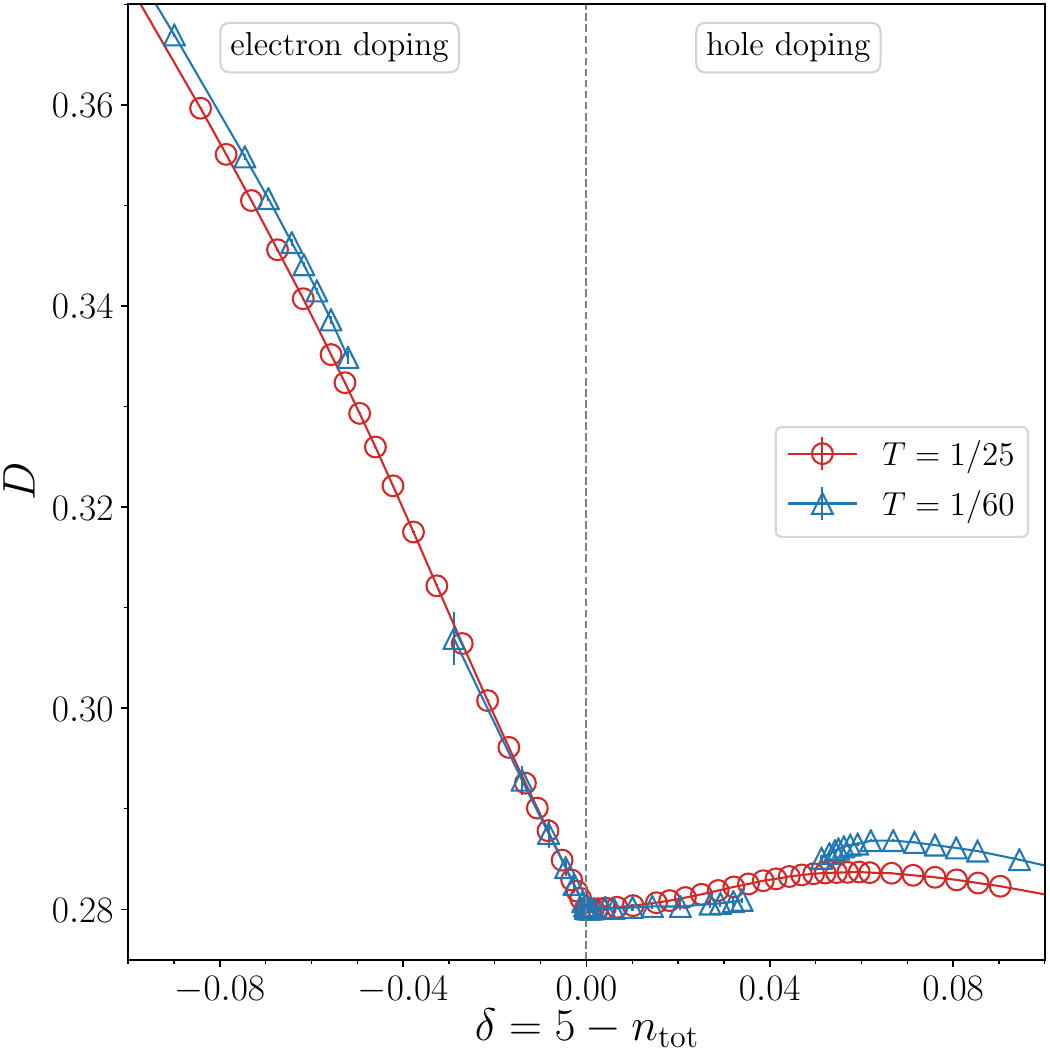}}
\caption{(a) Double occupancy $D$ of the copper orbital vs doping for different values of temperature. Left: electron doping, right: hole doping. Data are for the set of parameters of Eq.~\ref{eq:modelparameters} and $U_{d} =11.1$.
}
\label{fig:docc}
\end{figure}

This electron-hole asymmetry in the redistribution of the orbital $d$-$p$ character of the dopants arises from the Coulomb repulsion between electrons on the $3d$ orbitals. Therefore it is useful to investigate the double occupancy $D$ of the copper $3d$ orbital as a function of doping, as shown in Fig.~\ref{fig:docc}. 

There is a marked asymmetry in the behavior of $D(\delta)$. The variation of the double occupancy with doping is strongest upon electron doping. Upon electron doping, the double occupancy increases dramatically because the oxygen $2p$ orbitals are basically full and therefore the electrons go to doubly occupy the copper $3d$ orbitals. On the contrary, upon hole doping the variation of the double occupancy is smaller, because the doped holes mainly go to occupy the oxygen $2p$ orbitals.  
For either electron or hole doping, the double occupancy of the copper $3d$ orbital is smaller in the pseudogap than in the metal, implying that the pseudogap is more correlated than the metal. Furthermore, for either electron or hole doping, the temperature variation of the double occupancy of the copper $3d$ orbital has opposite behavior in the pseudogap than in the metal: $dD/dT>0$ in the pseudogap and $dD/dT<0$ in the metal.

\section{Effect of the interaction strength $U_{d}$}
\label{sec:Ud}

Up until now, the value of the interaction strength $U_d$ remained fixed to $U_d=11.1$. In this section we study the effect of varying $U_d$ on the phase diagram, on the local density of states, and on the occupancy of the copper and oxygen orbitals. 

\subsection{$U_d-\delta$ phase diagram}

\begin{figure}[ht!]
\centering{
\includegraphics[width=1.\linewidth]{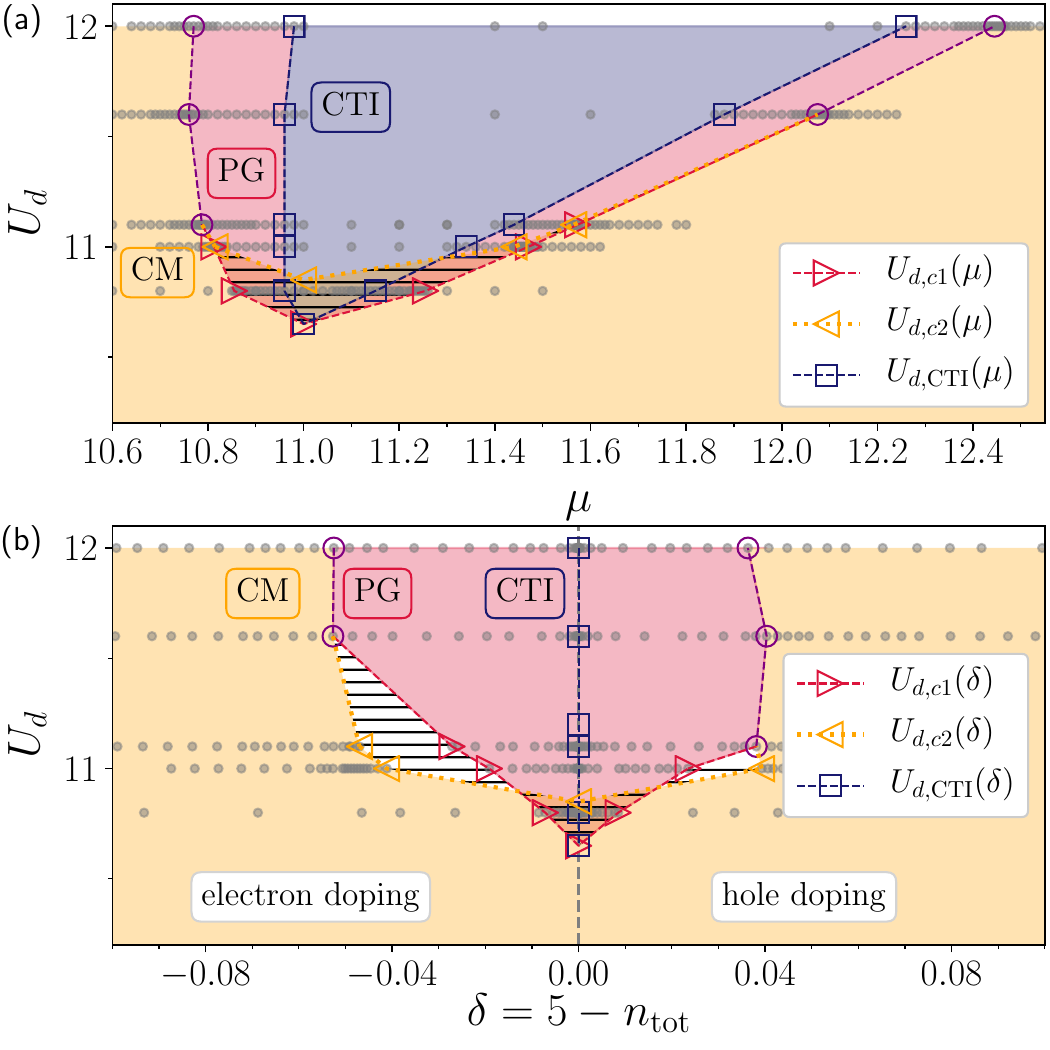}}
\caption{(a) $U_{d}$ vs $\mu$ phase diagram at $T=1/40$ and for the set of parameters of Eq.~\ref{eq:modelparameters}. Gray filled circles indicate the datapoints studied in our work. Charge-transfer insulator (CTI) phase is shown in dark blue, bounded by the dashed dark blue line with open squares. Emerging from this phase is the pseudogap (PG) phase shown in red, bounded by a dashed, red line with open red triangles. The correlated metal (CM) phase is shown in orange bounded by a dotted orange line with open orange triangles. Since the temperature of the critical end point decreases with increasing $U_d$, our scan of the $U_d-T-\mu$ space at constant temperature either cross the coexistence region (hatched) or the Widom line (open purple circles). 
(b) $U_{d}$ vs $\delta$ phase diagram at $T=1/40$. The CTI area of panel (a) collapses into a line at zero doping. The coexistence region of panel (a) maps into a region of coexistence (hatched with colored background) or into a forbidden region (hatched with white background). Left: electron doping, right: hole doping. 
}
\label{fig:phasedia2}
\end{figure}

Ref.~\cite{Lorenzo3band} showed that, for hole doping, the first order transition between the pseudogap and the metal is a first order surface in the $U_d-T-\delta$ space. Upon reducing $U_d$, this first order surface is continuously connected to the metal to charge-transfer insulator transition at zero doping. This first order surface ends at finite temperature in a critical line. This is in complete analogy with the behavior of the CDMFT solution of the single band 2D Hubbard model~\cite{sht, sht2}. 

Here, we extend the analysis of Ref.~\cite{Lorenzo3band} to both electron and hole dopings. To save computing time, we explored the $U_d-T-\delta$ space with a single cut at the fixed temperature $T =1/40$. Note that for $U_d=11.1$ this temperature is smaller than the temperature of the critical point of the pseudogap to metal transition for electron doping but is higher than the temperature of the critical point of the pseudogap transition for hole doping (see Fig.~\ref{fig:phasediagramT}). 
To explore the resulting $U_d-\delta$ plane, we calculate the isotherm $\delta(\mu)$ for different values of the interaction strength $U_d$, as we did in Sec.~\ref{sec:occupation} for $U_d=11.1$. 
The resulting phase diagram in the $U_d-\mu$ plane, or equivalently in the $U_d-\delta$ plane, is shown in Fig.~\ref{fig:phasedia2} [panels (a) and (b), respectively]. 

By tracking the plateau at $\delta(\mu)=0$ for different values of $U_d$, one obtains the V-shaped region of the charge-transfer insulator (blue region bounded by squares) in the $U_d-\mu$ phase diagram of Fig.~\ref{fig:phasedia2}(a). The charge-transfer insulator region in the $U_d-\mu$ plane collapses into a vertical line at $\delta=0$ in the $U_d-\delta$ plane of Fig.~\ref{fig:phasedia2}(b). 

By tracking the jumps in the isotherm $\delta(\mu)$ for different values of $U_d$, one obtains the coexistence region associated to the first order transition (hatched region bounded by right and left triangles) in the $U_d-\mu$ plane of Fig.~\ref{fig:phasedia2}(a). Note that the coexistence region can occur between a metal (orange region) and a charge-transfer insulator (blue region) or between a pseudogap (red region) and a metal, depending on the value of $U_d$. The coexistence region between two CDMFT solutions in the $U_d-\mu$ plane of Fig.~\ref{fig:phasedia2}(a) translates into regions where two solutions or no solution exist in the $U_d-\delta$ plane of Fig.~\ref{fig:phasedia2}(b) (hatched areas with colored and white background, respectively).

As shown in the temperature versus doping phase diagram or Fig.~\ref{fig:phasediagramT}, the first order transition for both electron and hole doping terminates in a critical point from which a supercritical crossover -- the Widom line -- emerges. Therefore we can also track the maxima of the charge compressibility which are a high temperature precursor of the first order transition (open purple circles). Since we are taking a cut of $U_d-T-\delta$ space at one fixed temperature, our data in Fig.~\ref{fig:phasedia2} indicate that the temperature of the second order critical line in the $U_d -T- \delta$ space decreases with increasing $U_d$. The decrease of the critical line with temperature is asymmetric between electron and hole doping, with a faster decrease upon hole doping. 

To summarise, Fig.~\ref{fig:phasedia2} indicates that the $U_d-T-\delta$ diagram hosts a surface of first order transitions. This transition surface progressively extends from zero doping to finite doping upon increasing $U_d$. This transition terminates in a second order critical line from which a surface of supercritical crossover emerges. This finding extends to electron doping the results of Ref.~\cite{Lorenzo3band} which focused on hole doping only. 

The phase diagram in Fig.~\ref{fig:phasedia2} enable us to gain a further understanding of the microscopic mechanism shaping the doping-driven metal to charge-transfer insulator transition. 
First, the $U_d-T-\delta$ phase diagram does not show a qualitative asymmetry upon electron or hole doping. A first order transition crosses the $\delta=0$ plane and extends on either doping sides. Hence, this first order transition is a robust feature against electron and hole dopings. A particle-hole symmetry only arises in the temperature and doping position of this transition. The pseudogap to metal transition at either particle and hole doping (see $T-\delta$ phase diagram for $U_d=11.1$ in Fig.~\ref{fig:phasediagramT}) are indeed connected in the $U_d-T-\delta$ plane, indicating their common origin in the Mott physics in two dimensions. 
Second, Fig.~\ref{fig:phasedia2} shows that the pseudogap phase emerges only above a threshold value of $U_d$ for opening a charge-transfer insulator at $\delta=0$, i.e. for $U>U_{d, {\rm MIT}}$, upon either electron or hole doping. This again suggests the origin of the pseudogap in the Mott physics in two dimensions, and that the pseudogap is a robust feature against electron or hole doping.

\subsection{Local density of states and spectral function at the Fermi level}

\begin{figure}[ht!]
\centering{
\includegraphics[width=1.\linewidth]{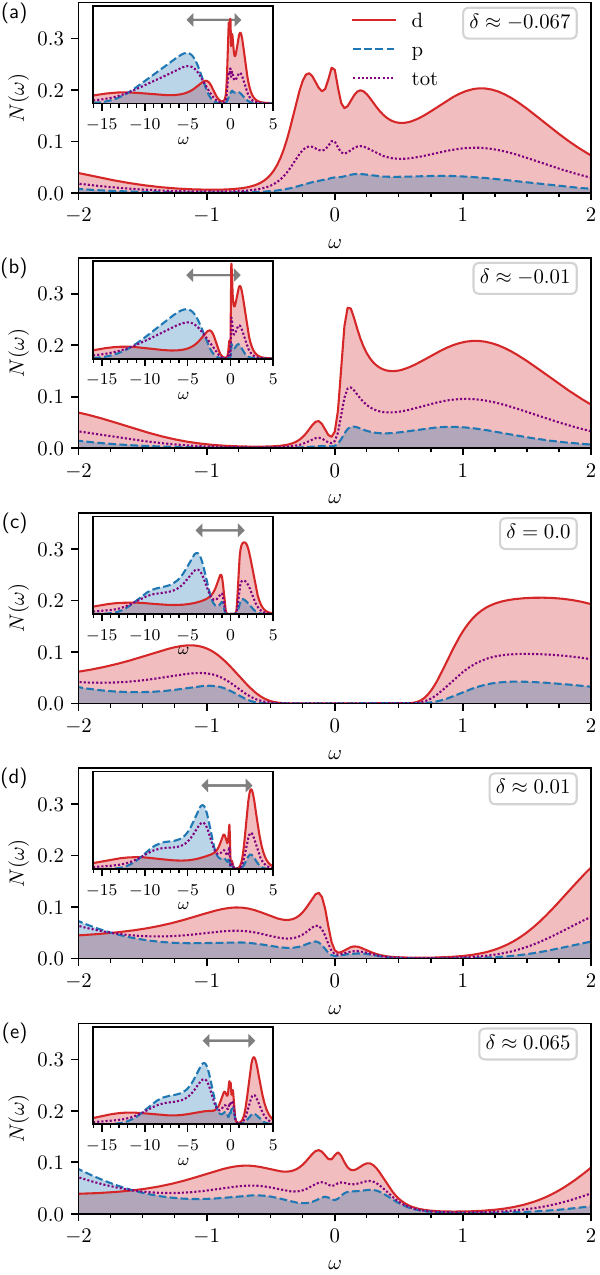}}
\caption{Projected local density of states $N(\omega)$ on the Cu $3d$ orbital (red solid line) and on the degenerate O $2p_x, 2p_y$ orbitals (blue dashed line), as well as the total density of states $N_{\rm tot}(\omega)=(N_d(\omega) +2N_p(\omega))/3$ (purple dotted line), for several doping levels. Each DOS is normalised to 1. 
(a),(b): electron doping; (c) charge-transfer insulator; (d),(e): hole doping. Data are for the set of parameters of Eq.~\ref{eq:modelparameters}, $U_{d} =12$ and $T=1/40$. The grey double arrow in the insets marks the distance $\Delta_g$ between the charge-transfer band and the upper Hubbard band on $N_{\rm tot}(\omega)$. The doping evolution of $\Delta_g$ is shown in Fig.~\ref{fig:dosPG2}. 
}
\label{fig:dos2}
\end{figure}
\begin{figure}[ht!]
\centering{
\includegraphics[width=1.\linewidth]{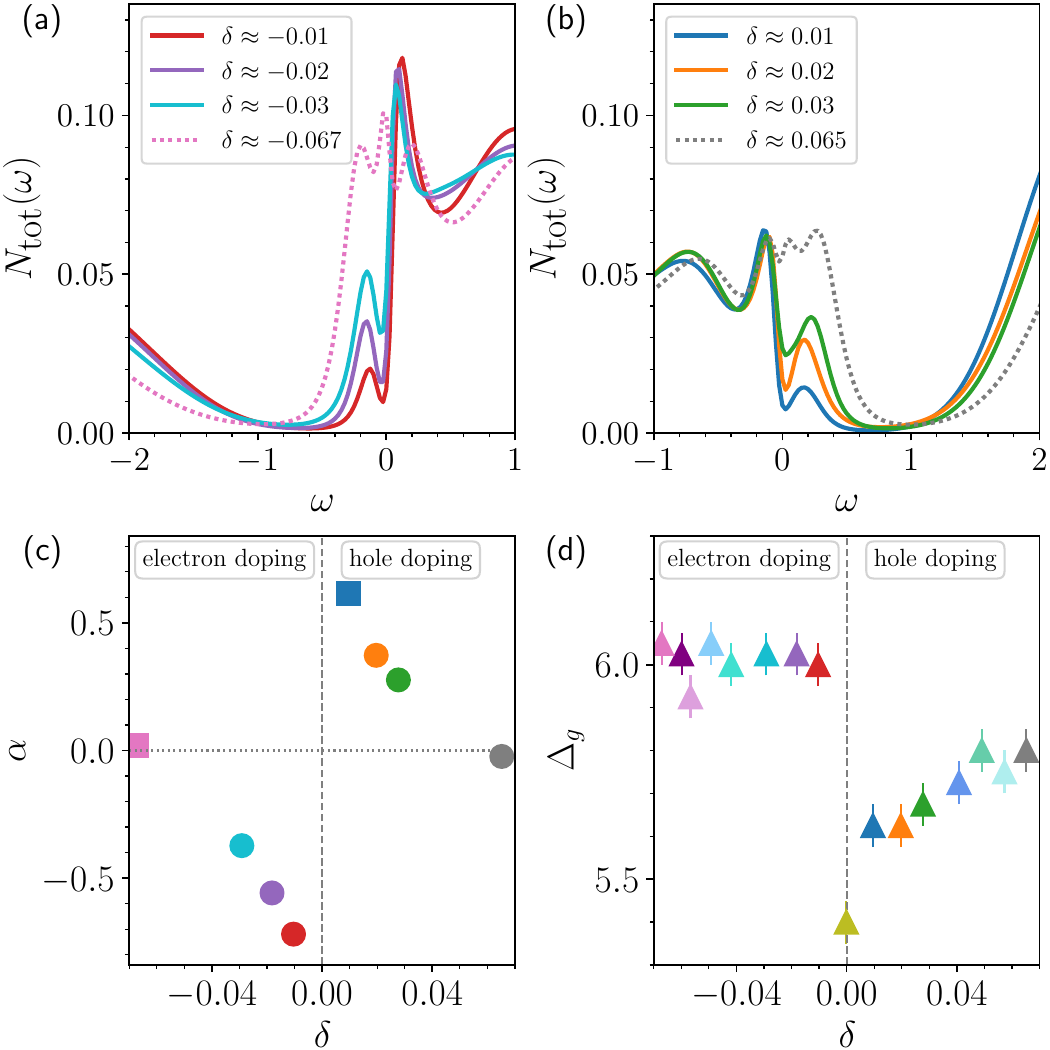}
}
\caption{Same as Fig.~\ref{fig:dosPG}, but for $U_d=12$. 
}
\label{fig:dosPG2}
\end{figure}
\begin{figure*}[ht!]
\centering{
\includegraphics[width=1.\linewidth]{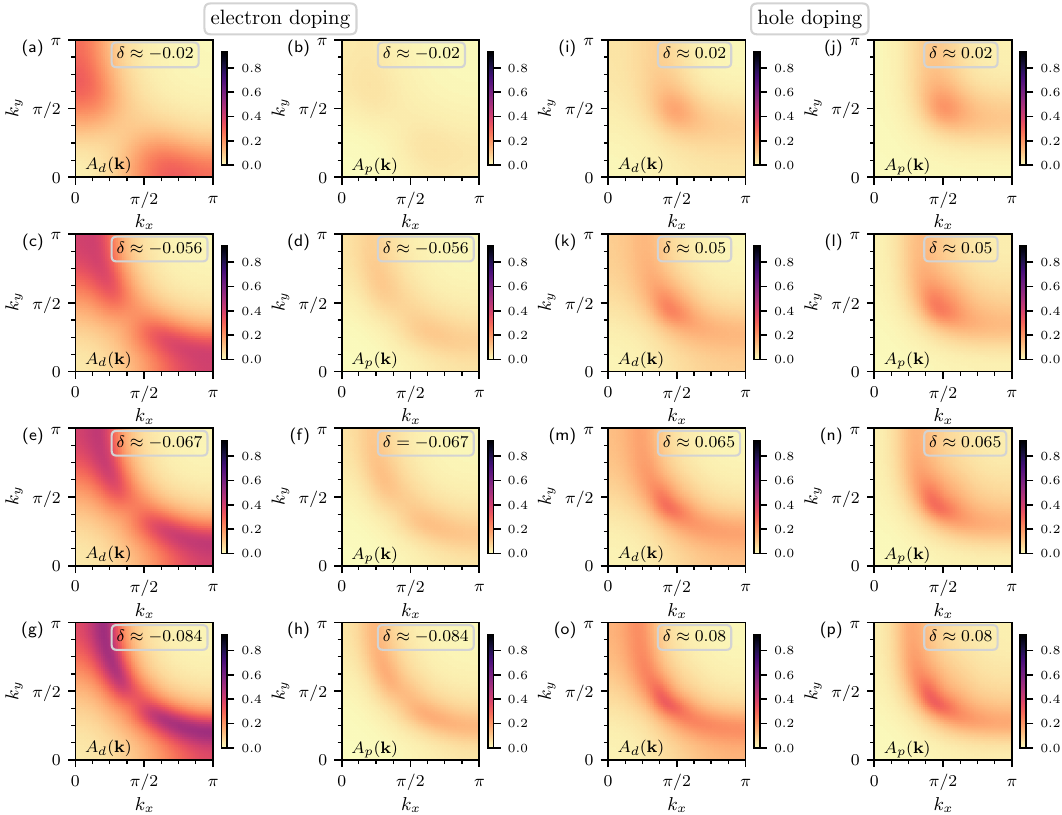}
}
\caption{Same as Fig.~\ref{fig:Akw1}, but for $U_d=12$ and $T=1/40$. 
}
\label{fig:Akw2}
\end{figure*}

Next we turn to the effect of the interaction strength $U_d$ on the density of states. Fig.~\ref{fig:dos2} shows the local density of states $N_d(\omega), N_{p}(\omega)$ and $N_{\rm tot}(\omega)$ at different values of doping for $U_d=12$, which is larger than the value $U_d=11.1$ considered in Fig.~\ref{fig:dos}. 

At zero doping [Fig.~\ref{fig:dos2}(c)], the charge-transfer gap is wider at $U_d=12$ than at $U_d=11.1$, as expected. 
At small electron doping [Fig.~\ref{fig:dos2}(b)], the pseudogap features at $U_d=12$ are more separated in energy from the charge-transfer band  than for $U_d=11.1$. 
Similarly, at small hole doping [Fig.~\ref{fig:dos2}(d)], the pseudogap features are more separated in energy from the upper Hubbard band than for $U_d=11.1$. 

The peaks forming the pseudogap about the Fermi level in the density of states for $U_d=12$ have a more dramatic height difference compared to $U_d=11.1$. As a result, the density of states at $U_d=12$ appears more asymmetric about the Fermi level, revealing that this asymmetry is an effect of Mott physics. 

Fig.~\ref{fig:dosPG2} analyses the low frequency electron-hole asymmetry of the pseudogap. As discussed in Fig.~\ref{fig:dos}, (i) the pseudogap fills in with increasing doping in absolute value (Fig.~\ref{fig:dosPG2}(a,b)), (ii) the pseudogap asymmetry $\alpha$ changes sign upon reversing the sign of the carriers (Fig.~\ref{fig:dosPG2}(c)), and (iii) the transfer of spectral weight giving rise to the pseudogap does not alter much the distance between the charge-transfer band and the upper Hubbard band (Fig.~\ref{fig:dosPG2}(d)).

Fig.~\ref{fig:Akw2} shows the intensity plots of the spectral functions $A_d({\bf k})$ and $A_p({\bf k})$ at the Fermi level at different values of doping, for $U_d=12$ and $T=1/40$. Overall, the spectral intensity is smaller at $U_d=12$ than at $U_d=11.1$, as expected from the increased correlation strength. As discussed in Fig.~\ref{fig:Akw1}, the pseudogap is characterized by a strong ${\bf k}$ dependent spectral functions [first row], with spectral intensity concentrated along the zone boundary at electron doping and along the zone diagonal at hole doping. Note however, that the features of the spectral intensity evolve with doping more gradually in Fig.~\ref{fig:Akw2} than in Fig.~\ref{fig:Akw1}: this is because in Fig.~\ref{fig:Akw2} the spectral intensity is shown at temperatures above the critical endpoint of the pseudogap to metal transition.

\subsection{Redistribution of the orbital character of the dopants}

\begin{figure}[t]
\centering{
\includegraphics[width=1.\linewidth]{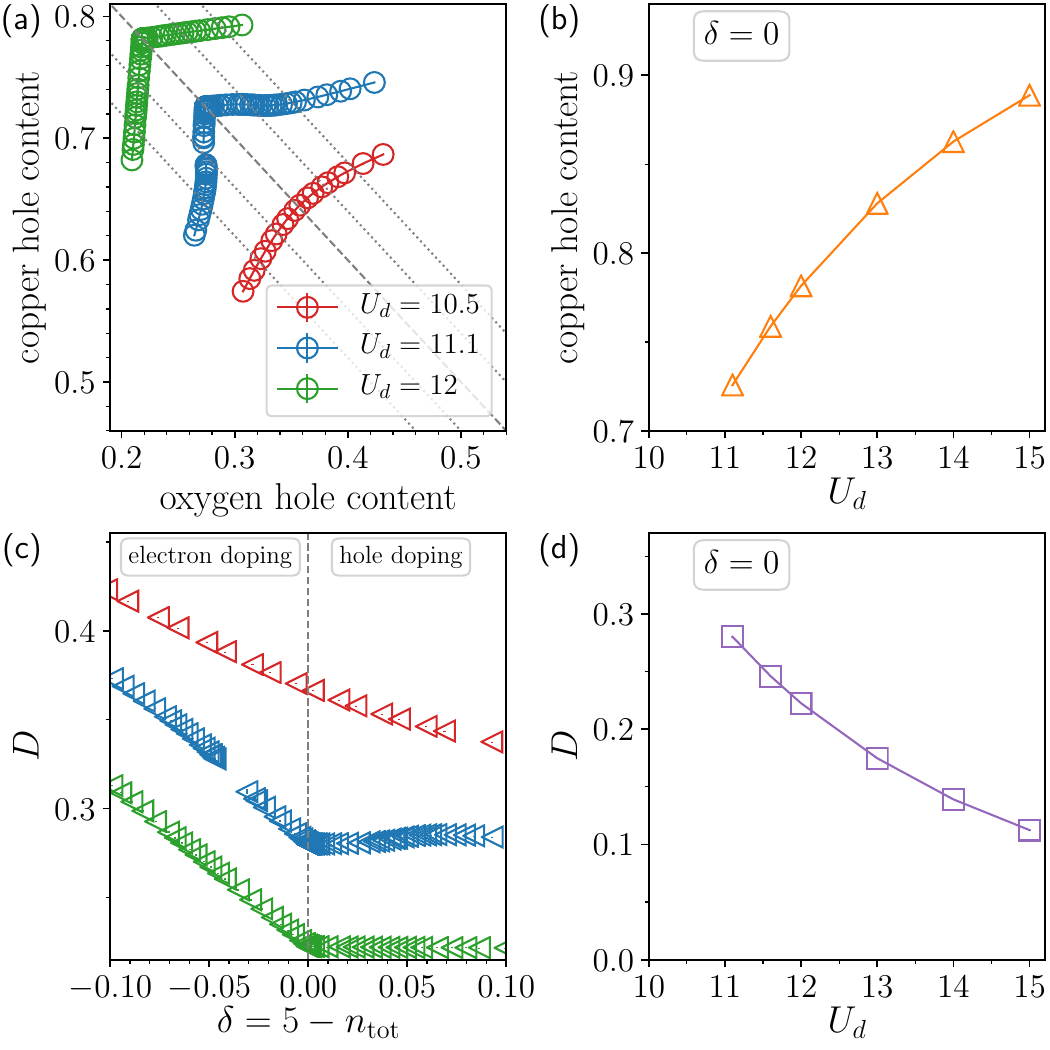}}
\caption{(a) Copper hole content vs oxygen hole content. Data are for the set of parameters of Eq.~\ref{eq:modelparameters} and different values of $U_d$. 
The grey dashed line marks the charge-transfer insulator at $\delta=0$. The grey dotted lines indicate constant electron or hole doping. 
(b) Copper hole content vs $U_{d}$ at $\delta=0$. (c) Double occupancy of the copper orbital vs doping for different values of $U_{d}$. (d) Double occupancy vs $U_{d}$ at $\delta=0$. Data in all panels are for $T=1/40$.
}
\label{fig:holecontent2}
\end{figure}

Next, we study how the interaction strength $U_d$ affects the occupancy of the Cu $3d$ and O $2p$ orbitals. Fig.~\ref{fig:holecontent2}(a) shows the copper hole content (i.e., $2-n_d$) versus oxygen hole content (i.e., $2(2-n_p)$) at $T=1/40$ and extends Fig.~\ref{fig:holecontent} to different values of $U_d$. 

A few trends appear, which give an indication about the mechanisms controlling the doping of a charge-transfer insulator. Overall, upon increasing $U_d$ the curves moves toward a large copper hole content and a small oxygen hole content (i.e., from bottom right to upper left region of the graph). 
This trend can be seen best along the $\delta=0$ line: Fig.~\ref{fig:holecontent2}(b) shows the increase of the Cu hole content at $\delta=0$ with increasing $U_d$. Since the Cu hole content is defined as $2-n_d$, this means that the occupation of the Cu $3d$ orbitals, $n_d$, decreases with increasing $U_d$. Equivalently, the mixed $d$-$p$ character of the hole in the charge-transfer insulator decreases with increasing $U_d$. 
Physically, decreasing the mixed $d$-$p$ character leads to an effective reduction of the hybridization between Cu and O orbitals. 
This is confirmed by the overall suppression of the double occupancy of the Cu $3d$ orbital with increasing $U_d$ (see Fig.~\ref{fig:holecontent2}(c),(d)). This is because the kinetic energy gain associated with the hopping between the Cu and O orbitals cannot overcome the large energy cost of doubly occupying the Cu $3d$ orbital. 

Next, let us turn to the slope of the curves in Fig.~\ref{fig:holecontent2}(a) close to their intersection point on the $\delta=0$ line. The overall shape of the curves in Fig.~\ref{fig:holecontent2}(a) is sensitive to the threshold value of $U_d$ for opening a charge-transfer insulator at $\delta=0$, namely $U_{d, \rm MIT}$. Below this critical threshold (see curve for $U_d=10.5$), the curves smoothly cross the $\delta=0$ line, whereas above this critical threshold the curves show a discontinuous slope across the $\delta=0$ line. 

We can identify three trends in the slopes of the curves close the $\delta=0$ line. 
(i) Below $U_{d, \rm MIT}$, the slope of the curves is large and positive, for either electron or hole doping. 
(ii) Above $U_{d, \rm MIT}$ and for hole doping, the slope of the curves slightly increases with increasing $U_d$. Equivalently, the curves have large and positive slope for small oxygen hole content, and a small and almost flat slope for higher oxygen hole content. 
(iii) Correspondingly, above $U_{d, \rm MIT}$ and for electron doping, the slope of the curves becomes slightly less steep with increasing $U_d$. Equivalently, the curves have larger slopes for small oxygen hole content, and almost a vertical slope for high oxygen hole content.

\section{Discussion}
\label{sec:discussion}

Our results show that the mere introduction of holes or electrons into an `ionic' charge-transfer insulator {\it does} produce an asymmetry in the doping evolution of the orbital character of the dopants, but it is {\it not sufficient} to produce a qualitative asymmetry in the nature of the doping-driven transition in the normal state. Hence, we are led to conclude that by {\it merely} doping with holes or electrons a charge-transfer insulator, one {\it cannot} take into account the marked electron-hole asymmetry observed in the normal state $T-\delta$ phase diagram of real cuprates. 

Our results rule out that the source of the observed electron-hole asymmetry lies solely in doping with holes or electrons the same parent charge-transfer insulator, but they cannot identify the source of that asymmetry. 
Thus, let us discuss separately the main insights that we can derive from our work about the {\it normal state} of hole-doped cuprates, electron-doped cuprates, and the cuprates for which ambipolar doping (i.e., doping with either holes or electrons) is experimentally achievable.

\subsection{Hole-doped cuprates}

Many studies have shown that the basic phenomenology of real hole-doped cuprate materials is well captured by that of a hole-doped charge-transfer insulator realised by the Emery model, and, by integrating out the high energy degrees of freedom, of a doped Mott insulator realised by the single-band 2D Hubbard model~\cite{Anderson:1987, Alloul2013, AMJulich}.

Similarly to the normal state of hole-doped cuprates, our results shows a charge-transfer insulator at $\delta=0$, followed by a pseudogap region~\cite{Alloul:1989} whose mechanism is associated to Mott physics plus short-range correlations~\cite{AMJulich, kyung, kancharla, hauleDOPING, ssht}. The pseudogap found in our calculations shows a striking particle-hole asymmetric shape with respect to the Fermi energy (Figs.~\ref{fig:dos}(d), \ref{fig:dosPG}(b), \ref{fig:dos2}(d), \ref{fig:dosPG2}(b)), as observed in experiments~\cite{Kohsaka:Science2007}. 
The pseudogap shows a strongly momentum dependent spectral intensity (Figs.~\ref{fig:Akw1}(a,b,i,j), \ref{fig:Akw2}(a,b,i,j)), with weight concentrated along the zone diagonal, in favourable agreement with the Fermi arcs revealed by angle-resolved photoemission (ARPES) experiments~\cite{Damascelli:RMP2003, Sobota:RMP2021}. 
The pseudogap to metal transition is followed at high doping by a maximum of thermodynamic entropy vs doping, whose position is almost doping independent and extends up to high temperature (Fig.~\ref{fig:phasediagramT}), as found in experiments~\cite{LoramJPCS2001, Tallon:2004}. 
Using the same methodology as the present work, Ref.~\cite{Lorenzo3band} has further shown that, for hole doping, the pseudogap to metal transition is actually concealed by a superconducting dome, and that the onset temperature of the pseudogap, $T^*$, ends abruptly is in the middle of the superconducting region, in agreement with experimental results~\cite{Collignon_Badoux_Taillefer:2017, Olivier:PRB2018}. 

Our present work also shows that doped holes mainly enter the oxygen orbitals (Fig.~\ref{fig:holecontent}), as found in experiments~\cite{Tranquada:PRB1987, Bianconi:SSC1987, Nucker:PRB1989, Chen:PRL1991, Chen:PRL1992, Sakurai:Science2011, Jurkutat:PRB2014, Gauquelin2014}. It also shows that upon increasing $U_d$, the copper hole content increases and the oxygen hole content decreases (Fig.~\ref{fig:holecontent2}(a)). 
It is interesting to compare our results in Fig.~\ref{fig:holecontent2}(a) to the experimental findings of Refs.~\cite{Jurkutat:PRB2014, Rybicki:NatComm2016} for different families of cuprates. 
First, the parent compound of hole-doped cuprates has a ratio of copper/oxygen hole content ranging approximately from $0.5/0.5$ of a Tl-based cuprate family~\cite{Jurkutat:PRB2014, Rybicki:NatComm2016} to $0.82/0.18$ of LSCO~\cite{Jurkutat:PRB2014, Rybicki:NatComm2016}. Hence, our results in Fig.~\ref{fig:holecontent2}(a) at large value of $U_d$ seem to be more appropriate to describe the phenomenology realised by LSCO. 
Second, prior work~\cite{Nicolas:PNAS2021, Nicolas:Master} has identified an anticorrelation between the oxygen hole content and the size of charge-transfer gap (i.e., the larger the oxygen hole content, the smaller the size of the charge-transfer gap). Since $U_d$ scales with the size of the gap, our results capture this trend (i.e., the smaller the $U_d$, and hence the smaller the charge-transfer gap, the higher the oxygen hole content). 
Third, the experimental trends for the slope of the curves close to the $\delta=0$ line and for hole-doped materials are opposite to our findings: experimentally, the curves have a small slope for small oxygen hole content (e.g. LSCO~\cite{Jurkutat:PRB2014, Rybicki:NatComm2016}), and a large slope for higher oxygen content (e.g. Tl-based cuprate family~\cite{Jurkutat:PRB2014, Rybicki:NatComm2016}). This suggests that to describe this experimental trend, one cannot just simply reduce $U_d$ in the Emery model, but other effects, such as those produced by tuning the bare charge-transfer energy $|\epsilon_p - \epsilon_d|$, the hybridization $t_{pd}$, as well as the interaction strength $U_p$ on the oxygens should be considered. We leave this investigation for further work.

\subsection{Electron-doped cuprates}

Let us now turn to the electron-doped case. It is long known that the phenomenology of real electron-doped materials such as PCCO and NCCO differs from that of real hole-doped materials. Let us stress that our works focuses on the normal state CDMFT solution of the Emery model, i.e. we do not allow symmetry breaking states. Physically, this means that the normal state occurs at all temperatures.    
In sharp contrast to experiments, our work finds a qualitative symmetry in the nature of electron and hole doped side of the $T-\delta$ normal state phase diagram (see Fig.~\ref{fig:phasediagramT}).  Hence we are led to conclude that the charge-transfer insulator regime studied in our theory work is {\it not} the relevant regime for the parent compound of actual electron-doped cuprate materials such as PCCO and NCCO. This also implies that the strongly correlated pseudogap arising from Mott physics in 2D found in our work is {\it not} relevant for the the pseudogap found in electron-doped cuprates such as PCCO and NCCO~\cite{Armitage:PRL2001, Motoyama:2007}, which is associated not to Mott physics but to a fluctuating precursors of a long-range antiferromagnetic order~\cite{Vilk:1997}. For example, the doping evolution of the spectral function found in our calculations is not captured by ARPES experiments on NCCO~\cite{Damascelli:RMP2003, Sobota:RMP2021}. 

In addition, our findings about the doping redistribution of the orbital character of the dopant carriers correctly capture that doped electrons mainly enter the copper orbital (Fig.~\ref{fig:holecontent}), but the order of the ratio of copper/oxygen hole content is far away from the experimental results on NCCO and PCCO~\cite{Jurkutat:PRB2014, Rybicki:NatComm2016}. 
Indeed, the parent compound for these materials has a ratio of copper/oxygen hole content around $0.55/0.45$~\cite{Jurkutat:PRB2014, Rybicki:NatComm2016}, which is not captured by our set of parameters. 
This reinforces the idea that the charge-transfer insulator regime studied in our work is not the relevant regime for actual electron-doped cuprates NCCO and PCCO.

Second, the high oxygen hole content in electron-doped materials correlates with their smaller charge-transfer gap at zero doping~\cite{Jurkutat:PRB2014, Ohta:PRL1991}. This is compatible with prior work suggesting that electron-doped materials are less correlated than their hole-doped counterpart~\cite{tremblayR, st, Cedric:NatPhys2010, cedricApical}. 
Our finding that $U_d$ is one possible microscopic mechanism governing the redistribution of electrons and holes is compatible with this viewpoint. 
This is because (i) below $U_{d, {\rm MIT}}$ we found no strongly correlated pseudogap (see Fig.~\ref{fig:phasedia2}) and (ii) because decreasing $U_d$ leads to an increase of the oxygen hole content (see Fig.~\ref{fig:holecontent2}(a)). 
In that respect, our findings are compatible with studies~\cite{Cedric:NatPhys2010, cedricApical, st} indicating that actual electron-doped cuprates may lie on the metallic side of the metal to charge-transfer insulator boundary of the Zaanen-Sawatsky-Allen diagram.  Furthermore, studies~\cite{Chloe:PRB2024} on the Emery model set on the metallic side of the metal to charge-transfer insulator boundary show a better electronic screening and hence less electronic correlations with increasing electron doping. 
Note that our work focuses only on the effects of $U_d$. However other effects, such as those produced by tuning the bare charge-transfer energy $|\epsilon_p - \epsilon_d|$, can change the effective interaction strength, hence further work is needed to clarify the interplay of these effects. We leave this for further investigations. 

Finally, it is interesting to mention that an experimental advance in the annealing method, known as ``protect annealing'', has shown the possibility to suppress antiferromagnetic fluctuations and thus the antiferromagnetic pseudogap in the electron doped Pr$_{1.3-x}$La$_{0.7}$Ce$_x$CuO$_4$ (PLCCO) single crystals~\cite{Horio:NatCommun:2016, Lin:PRR:2021}. The suppression of fluctuating precursors of long range antiferromagnetic order could reduce the asymmetry between the temperature vs electron doping and temperature vs hole doping phase diagram of cuprates~\cite{Lin:PRR:2021}. Therefore the CDMFT normal state results could be relevant for protected annealed PCCO, as already pointed out in Ref.~\onlinecite{Sakai:JPSJ2023} in the context of single band 2D Hubbard model. 
Furthermore, Refs.~\cite{Horio:NatCommun:2016, Song:PRL2017, Lin:PRR:2021} pointed out a possible discrepancy between the nominal and actual doping concentrations in electron-doped cuprates: using the Luttinger number as the doping axis, one obtains a $T-\delta$ phase diagram of electron-doped cuprates, in which the doping range of the antiferromagnetic and superconducting phases are similar to that of hole-doped cuprates.

 \subsection{The ambipolar Sr$_{1-x}$La$_{x}$CuO$_{2+y}$ cuprate films}
 \label{sec:SLCO}
 
As pointed out in the Introduction, the ambipolar doping (i.e., doping with either electrons or holes) of a given cuprate material is experimentally difficult~\cite{ift}. However, recent technical advances have enabled doping with either electrons or holes a few cuprate systems~\cite{Segawa:PRB2006, segawa:NatPhys2010, Zhong:PRL2020}. 
The experimental work on ambipolar doping the cuprate films of Sr$_{1-x}$La$_{x}$CuO$_{2+y}$ (SLCO)~\cite{Zhong:PRL2020} seems particularly relevant for our findings. 

First, using scanning transmission electron microscopy (STEM), Ref.~\cite{Zhong:PRL2020} shows that the high-energy features of the SLCO bandstructure remains essentially {\it symmetric} upon either electron or hole doping, despite the location of the doped carriers (doped holes mainly enter into the oxygens, doped electrons mainly enter into the copper). In particular, the charge-transfer gap remains essentially constant (around $1.3$~eV), regardless the hole vs electron nature of the carriers.  
This echoes our findings that (i) the distance between the charge-transfer band and the upper Hubbard band does not vary much with doping (see Fig.~\ref{fig:dosPG}(d), \ref{fig:dosPG2}(d)) and (ii) ambipolar doping of a well developed (i.e. an `ionic') charge-transfer gap, no qualitative asymmetry arises in the nature of the doping-driven transition in the normal state (see Fig.~\ref{fig:phasediagramT}). Prior mean-field calculations~\cite{Kotiar:PhysC1988} also found an intriguing electron-hole symmetry in the bandstructure of a doped charge-transfer insulator, irrespective of the location of the doped carriers. 

Second, our findings suggest that doping of an `ionic' charge-transfer gap should give rise to a strongly correlated pseudogap -- i.e. a pseudogap associated purely with Mott physics in 2D --  upon either electron or hole doping (see Fig.~\ref{fig:phasediagramT}). Intriguingly, the STEM spectra of SLCO cuprate films also display pseudogap features upon either electron or hole doping (see supplemental material of Ref.~\cite{Zhong:PRL2020}). Furthermore, and similar to our results (see Figs.~\ref{fig:dos}, \ref{fig:dosPG}(a,b), \ref{fig:dos2}, \ref{fig:dosPG2}(a,b)), the STEM spectra show a pseudogap with a two-peak structure characterised by a marked asymmetry about the Fermi energy, for either electron or hole doping.  Both in experiments and in our results, the electron-hole asymmetry of the pseudogap around the Fermi energy depends on the nature of the dopant carriers: for hole doping the pseudogap has a more pronounced peak below the Fermi energy, whereas for electron doping the pseudogap has a more pronounced peak above the Fermi energy. Therefore, upon changing between electron and hole doping, the electron-hole asymmetry of the pseudogap is reversed (see Figs.~\ref{fig:dosPG}(c), \ref{fig:dosPG2}(c)). Hence our work provides a possible theoretical framework to explain the experimental results of  Ref.~\cite{Zhong:PRL2020}. 

Although the qualitative comparison of our findings with the experimental results of Ref.~\cite{Zhong:PRL2020} seems promising, it nevertheless needs to be interpreted with caution. Experimentally, Ref.~\cite{Zhong:PRL2020} found that upon varying doping, the crystal structure remain unchanged, but the out-of-plane lattice constant shows a small change (about $0.2$~\AA), due to the intake of the apical oxygens. As a result, the doped holes enter mostly the apical oxygens, which are not taken into account in our work. Furthermore, both in our theory work and in Ref.~\cite{Zhong:PRL2020}, the interplay of the pseudogap with ordered states such as superconductivity and antiferromagnetism still deserve further study.

\section{Conclusions}
\label{sec:conclusions}

In this work, we have considered similarities and differences of doping with electrons and holes a charge-transfer insulator realised by the Emery model in the normal state, solved with CDMFT. We have addressed two main questions: (i) what is the nature of the doping-driven metal to charge-transfer insulator transition and (ii) how the orbital character of the dopant carriers evolves with either doping paths. This work expands Ref.~\cite{Lorenzo3band}, which has used the same methodology of this work and has focused on the hole-doping case only. 

Regarding the first question about the nature of the metal to charge-transfer insulator transition, we find a purely electronic two-stage doping-driven transition upon either electron or hole doping, i.e. irrespective of the orbital character of the dopants.  Above a critical threshold of $U_d$ for opening a charge-transfer gap at $\delta=0$, there is a continuous transition between a charge-transfer insulator and a strongly correlated pseudogap at $\delta=0$, followed by a discontinuous transition between a strongly correlated pseudogap and a correlated metal at finite doping. The pseudogap to metal transition ends in a critical point at finite temperature, from which emerges a supercritical crossover marking an anomalous enhancement in the charge compressibility. This two-stage electronic transition occurs upon either electron or hole doping (see $T-\delta$ phase diagram in Fig.~\ref{fig:phasediagramT}). By varying the strength of the electron-electron interaction $U_d$ on the copper orbital, we find that the pseudogap to metal transition on either doping path is connected to the metal to charge-transfer insulator transition at zero doping (see $U_d-\delta$ phase diagram in Fig.~\ref{fig:phasedia2}). Thus, a surface of first order transition in the $U_d-T-\delta$ space crosses zero doping and progressively extends to larger dopings with increasing $U_d$. 

From these findings we identify three main conclusions.  (i) There is no qualitative electron-hole asymmetry in the nature of the doping-driven transition of a charge-transfer insulator in the normal state. A quantitative electron-doping asymmetry is however found on the doping level at which this transition occurs and on the temperature at which this transition ends. 
(ii) In agreement with the results of Ref.~\cite{Lorenzo3band}, the pseudogap is a strongly correlated phase that only emerges for $U_d$ larger than the critical threshold to open a charge-transfer insulator at $\delta=0$ (one hole per CuO$_2$ unit cell), hence the name `strongly correlated pseudogap'.  
(iii) Beyond the results of Ref.~\cite{Lorenzo3band}, the pseudogap {\it on either doping path} has a common origin in Mott physics in two dimensions, that is, it is a robust phenomenon emerging from a doped correlated insulator, regardless of the (hole vs electron) nature of the dopants. 

Regarding the second question of our work, namely the doping evolution of the orbital character of the dopant carriers, we find a striking electron-hole asymmetry. Doped electrons predominantly go to occupy the copper orbitals, whereas doped holes mainly enter the oxygen orbitals (see Fig.~\ref{fig:holecontent}). Furthermore, upon increasing the electron-electron interaction $U_d$ on the copper orbitals, the copper hole content increases and the oxygen hole content decreases (see Fig.~\ref{fig:holecontent2}). This allows us to identify the interaction on the copper orbitals as one possible microscopic mechanism governing the redistribution of electrons and holes.

Taken together, our results show that ambipolar doping of a charge-transfer insulator produces an asymmetry in the orbital character of the dopants, but it is not sufficient to produce an asymmetry in the qualitative features of the normal state doping-driven transition. Hence, our findings rule out that the origin of the observed normal state electron-hole asymmetry in cuprates lies solely in doping with electrons and holes a charge-transfer insulator. In the Discussion Sec.~\ref{sec:discussion}, we commented of the implications for actual hole- and electron-doped cuprates. 

Our work paves the way for future investigations of the Emery model in other parameter regimes. In particular, our work focuses on a clear (i.e., `ionic') charge-transfer regime. It would be interesting to investigate how our results are modified by setting the Emery model in the intermediate (or `covalent') regime in the Zaanen-Sawatzky-Allen scheme~\cite{zsa}.

\begin{acknowledgments}
This work has been supported by the Canada First Research Excellence Fund. Simulations were performed on computers provided by the Canadian Foundation for Innovation, the Minist\`ere de l'\'Education des Loisirs et du Sport (Qu\'ebec), Calcul Qu\'ebec, and Digital Research Alliance of Canada.
\end{acknowledgments}


%

\end{document}